\documentclass[review,12pt]{elsarticle}

\usepackage{amssymb}
\usepackage{amsthm}
\usepackage{framed} 
\usepackage{amsmath,color}
\usepackage{mathrsfs}
\usepackage{graphicx}
\usepackage{epstopdf}
\usepackage{float}
\usepackage{caption}
\usepackage{subcaption}
\usepackage{bm}
\usepackage{bbm}
\usepackage{mathrsfs}
\usepackage{cleveref}
\usepackage{soul}
\usepackage{accents}
\usepackage{color,soul} 
\usepackage{color} 
\usepackage{bm}
\usepackage{multirow} 
\usepackage[margin=1.78cm]{geometry}
\biboptions{sort&compress}
\soulregister\citep7 
\soulregister\citet7 
\soulregister\citealp7 
\newsavebox{\measurebox} 
\usepackage{titlesec} 
\usepackage[T1]{fontenc} 
\usepackage{lmodern} 
\input{glyphtounicode}  

\def\onedot{$\mathsurround0pt\ldotp$}
\def\cdddot#1{
  \mathbin{\vcenter{\baselineskip.67ex
    \hbox{\onedot}\hbox{\onedot}\hbox{\onedot}%
  }}%
}

\journal{ }

\makeatletter
\def\@author#1{\g@addto@macro\elsauthors{\normalsize%
    \def\baselinestretch{1}%
    \upshape\authorsep#1\unskip\textsuperscript{%
      \ifx\@fnmark\@empty\else\unskip\sep\@fnmark\let\sep=,\fi
      \ifx\@corref\@empty\else\unskip\sep\@corref\let\sep=,\fi
      }%
    \def\authorsep{\unskip,\space}%
    \global\let\@fnmark\@empty
    \global\let\@corref\@empty  
    \global\let\sep\@empty}%
    \@eadauthor={#1}
}
\makeatother

\titleformat{\paragraph}
{\normalfont\normalsize\itshape}{\theparagraph}{1em}{}
\titlespacing*{\paragraph}
{0pt}{3.25ex plus 1ex minus .2ex}{1.5ex plus .2ex}

\begin{document}

\begin{frontmatter}



\title{Griffith-based analysis of crack initiation location in a Brazilian test}


\author{Yousef Navidtehrani\fnref{Uniovi}}
\author{Covadonga Beteg\'{o}n \fnref{Uniovi}}
\author{Robert W. Zimmerman \fnref{ICearth}} 
\author{Emilio Mart\'{\i}nez-Pa\~neda\corref{cor1}\fnref{IC}}
\ead{e.martinez-paneda@imperial.ac.uk}

\address[Uniovi]{Department of Construction and Manufacturing Engineering, University of Oviedo, Gij\'{o}n 33203, Spain}

\address[ICearth]{Department of Earth Science and Engineering, Imperial College London, London SW7 2AZ, UK}

\address[IC]{Department of Civil and Environmental Engineering, Imperial College London, London SW7 2AZ, UK}

\cortext[cor1]{Corresponding author.}

\begin{abstract}
The Brazilian test has been extremely popular while
prompting significant debate. The main source of controversy is rooted in its \emph{indirect} nature; the material tensile strength is inferred upon assuming that cracking initiates at the centre of the sample. Here, we use the Griffith criterion and finite element analysis to map the conditions (jaws geometry and material properties) that result in the nucleation of a centre crack. Unlike previous studies, we do not restrict ourselves to evaluating the stress state at the disk centre; the failure envelope of the generalised Griffith criterion is used to establish the crack nucleation location. We find that the range of conditions where the Brazilian test is valid is much narrower than previously assumed, with current practices and standards being inappropriate for a wide range of rock-like materials. The results obtained are used to develop a protocol that experimentalists can follow to obtain a valid estimate of the material tensile strength. This is showcased with specific case studies and examples of valid and invalid tests from the literature. Furthermore, the uptake of this protocol is facilitated by providing a MATLAB App that determines the validity of the experiment for arbitrary test conditions. \\
\end{abstract}


\end{frontmatter}



\section{Introduction}
\label{Sec:Introduction}

\begin{figure}[H]
    \centering
    \begin{subfigure}[t]{0.35\textwidth}
    \includegraphics[width=\textwidth]{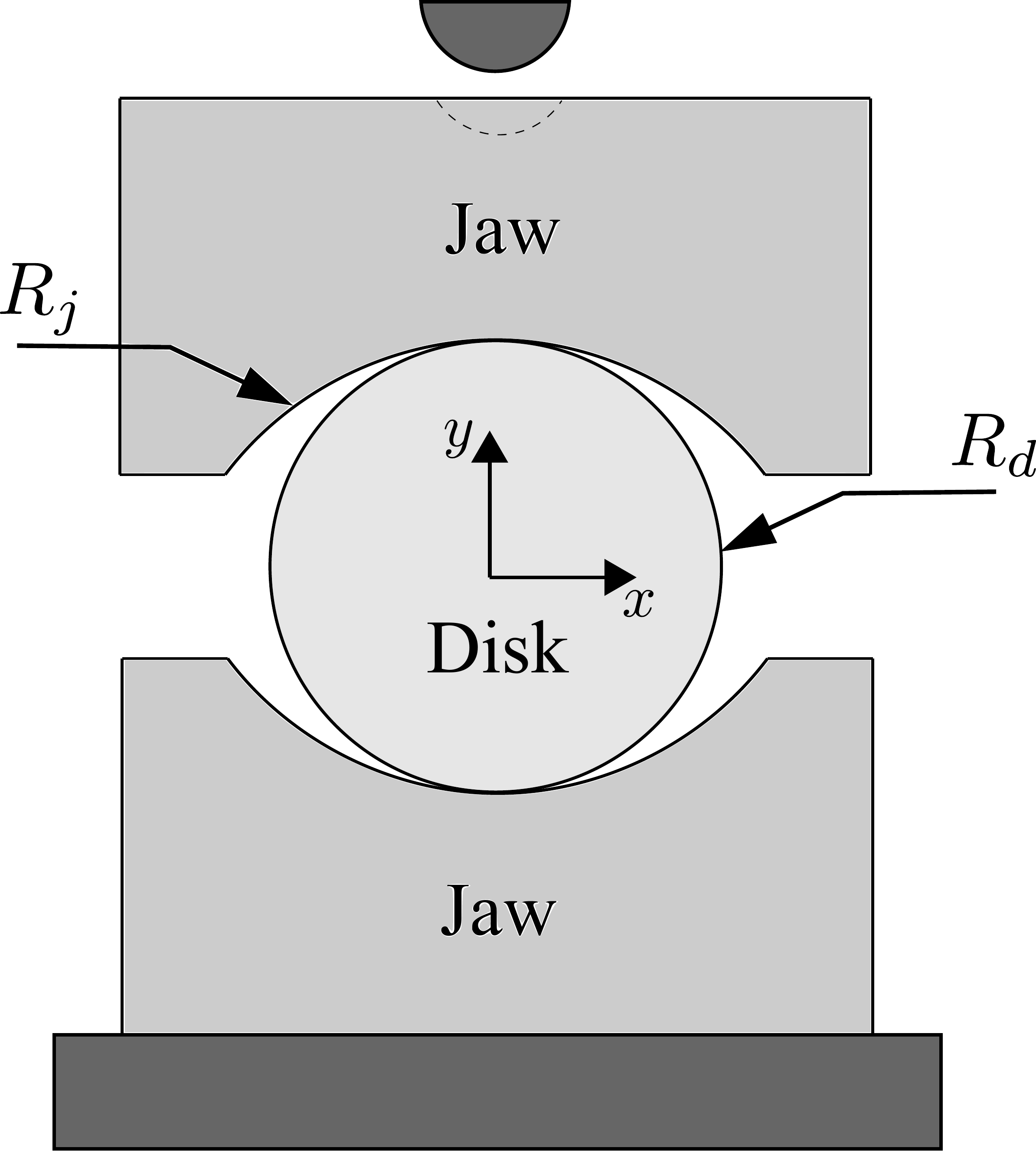}
    \caption{}
    \label{}
    \end{subfigure}\hspace{1 cm}  
    \begin{subfigure}[t]{0.305\textwidth} 
    \includegraphics[width=\textwidth]{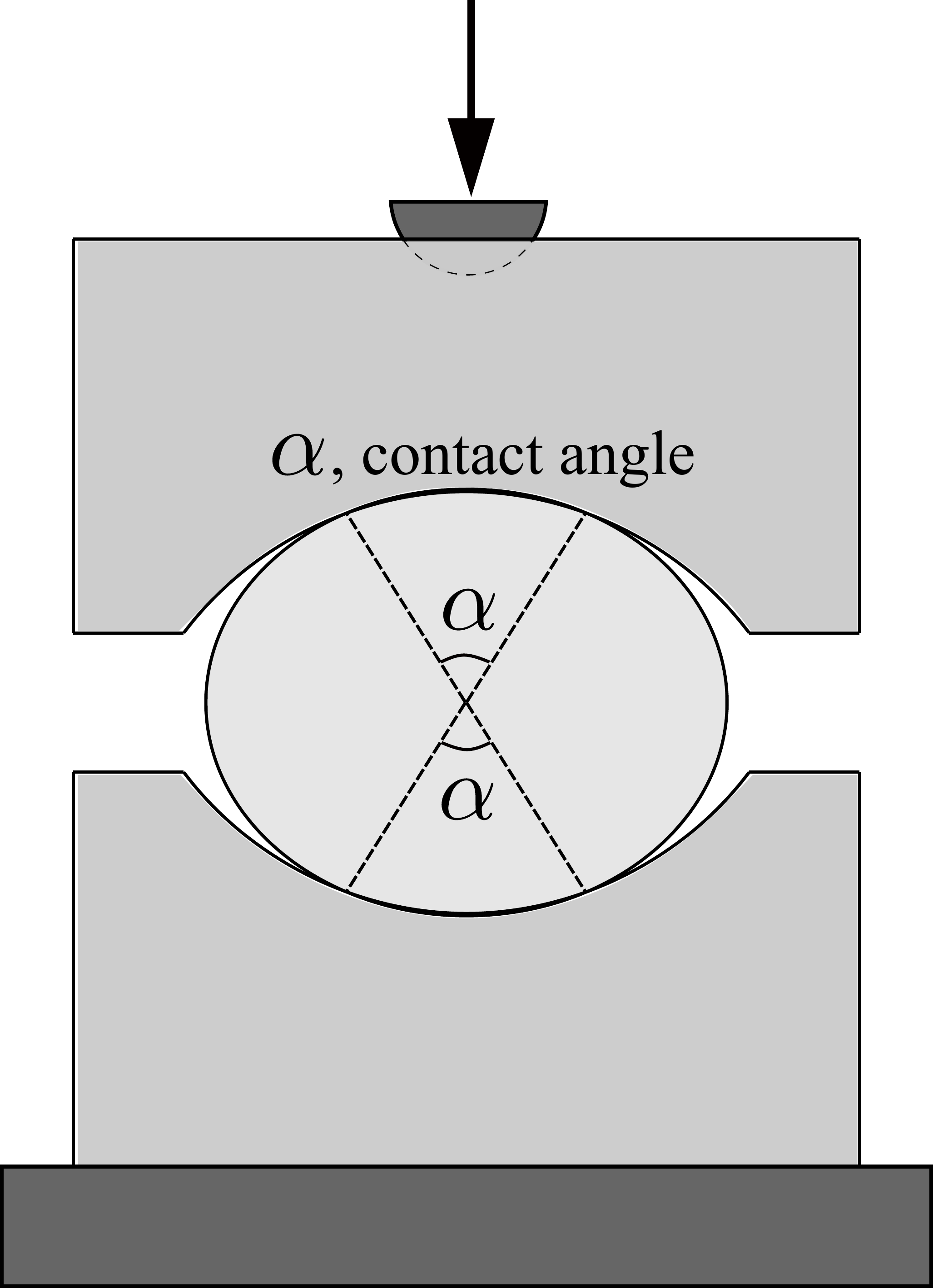}
    \caption{}
    \label{}
    \end{subfigure}
    \caption{Brazilian test configuration in the (a) undeformed, and (b) deformed states. The sketch shows the main variables: the jaw radius $R_j$, the disk radius $R_d$, and the contact angle at failure $\alpha$. A reaction force $P$ is measured.}
    \label{fig:BrazilianSketch1}
\end{figure}

The Brazilian test, also known as the Splitting Tensile Strength test, is arguably the most popular laboratory experiment for estimating the tensile strength of rocks and other quasi-brittle materials \cite{Li2013}. It was, independently, first proposed by Carneiro \cite{Carneiro1943} and Akazawa \cite{Akazawa1943} in 1943, and has been considered a standardised test since 1978, when it was included as a Suggested Method of the International Society for Rock Mechanics (ISRM) \cite{Bieniawski1978}. 
As shown in Fig. \ref{fig:BrazilianSketch1}, the test is comprised of two loading jaws, typically made of steel, and a disc-shaped sample. The jaws are configured so as to contact the sample at diametrically-opposed surfaces. Critical variables are the jaw radius, $R_j$, the disk radius, $R_d$, the disk thickness $t$, the measured reaction force $P$, and the contact angle $\alpha$. 

Assuming isotropic, linear elastic material behaviour, Hondros \cite{Hondros1959} derived an equation that relates the measured load $P$ and contact angle $\alpha$ with the maximum principal stress at the centre of the disk:
\begin{equation}\label{eq:Hondros2}
 \left( \sigma_1 \right)_{x=0,y=0} =\frac{2 P }{\pi R_d t \alpha}\left(\sin \alpha- \frac{\alpha}{2}\right) \, .
\end{equation}

Thus, from the critical values of $P$ and $\alpha$ at failure, one can use Eq. (\ref{eq:Hondros2}) to estimate the material tensile strength $\sigma_t$ upon assuming that the maximum value of $\sigma_1$ is attained at the centre of the disk: $\sigma_t = \left( \sigma_1 \right)_{x=0,y=0}$. However, Eq. (\ref{eq:Hondros2}) is derived assuming the application of a uniform pressure. Moreover, being able to experimentally measure the contact angle at failure is far from trivial. Consequently, standards are built upon the assumption of a zero contact angle, simplifying Eq. (\ref{eq:Hondros2}) to the case of a concentrated load:
\begin{equation}\label{eq:S1withLoad}
    \left( \sigma_1 \right)_{x=0,y=0} = \frac{P}{\pi R_d t} \, , \,\,\,\,\,\,\, \text{for} \,\,\,\,\,\,\, \alpha \to 0 \, .
\end{equation}

\noindent Eq. (\ref{eq:S1withLoad}) is often referred to as the Hondros's point load solution or the Hertz solution \cite{Timoshenko1951}. Using Eq. (\ref{eq:S1withLoad}), the material tensile strength can be readily estimated from the critical load ($P_c$): $\sigma_t = \left( \sigma_1 \right)_{x=0,y=0}=P_c/(\pi R_d t)$. However, this \emph{indirect} approach builds upon a number of assumptions; most notably: (i) the load is assumed to be a concentrated point load, and (ii) cracking initiates from the centre of the disk. In practice, fulfilling these two assumptions depends on the choices of test geometry and material. Numerical computations show the existence of three regimes. Sufficiently low contact angles will satisfy Eq. (\ref{eq:S1withLoad}) and lead to a maximum value of $\sigma_1$ at the disk centre. As the contact angle increases, Eq. (\ref{eq:S1withLoad}) is no longer satisfied, but the maximum magnitude of the tensile principal stress is still attained at the centre. And finally, if the contact angle is sufficiently large then not only is Eq. (\ref{eq:S1withLoad}) not satisfied but also the location of the maximum tensile stress moves away from the disk centre. Thus, the validity of the Brazilian test is sensitive to the contact angle at failure, which is itself dependent on the elastic properties of the disk and jaws (Young's moduli $E_d$, $E_j$; Poisson's ratios $\nu_d$, $\nu_j$), the sample and jaw radii ($R_d$, $R_j$), and the critical load (i.e., the material strength). Not surprisingly, this sensitivity to material and test parameters has fostered significant discussion in the academic literature. Despite the current popularity of the Brazilian test, early studies highlighted the sensitivity of the crack initiation location to the contact angle and questioned its use \cite{Fairhurst1964,Hudson1972}. The debate is very much open and a myriad of papers have been published trying to shed light on the validity regimes of the Brazilian test using theoretical, numerical and experimental tools. Recent examples include the work of Alvarez‑Fernandez and co-workers \cite{Alvarez-Fernandez2020}, who investigated, experimentally and analytically, the influence of the contact angle in the stress distribution and the failure load in slate. They reported that contact angles in the range $23-32^{\circ}$ were the most suitable to achieve crack initiation near the disk centre. Markides and Kourkoulis \cite{Markides2016} used analytical methods to evaluate the sensitivity of the stress state to the jaw's curvature, delimiting the conditions where Eq. (\ref{eq:S1withLoad}) is applicable. Gutierrez-Moizant \textit{et al.} \cite{Gutierrez-Moizant2020} conducted Brazilian tests in concrete with various contact angles and recommended using a loading arc of $20^\circ$. Bouali and Bouassida \cite{Bouali2021} investigated the role of the contact angle for both concrete and mortar, concluding that $20^\circ$ was the most suitable contact angle for concrete while $10^\circ$ was recommended for mortar. Garcia-Fernandez \textit{et al.} \cite{Garcia-Fernandez2018} conducted Brazilian tests in PMMA samples, which enabled them to visualise the crack initiation process and demonstrated the important role of the contact angle. Zhao and co-workers \cite{Zhao2021} used acoustic emission to investigate the role of the experiment setup on the crack nucleation event. Aliabadian \textit{et al.} \cite{Aliabadian2019} showed, using Digital Image Correlation (DIC), that the location of crack nucleation was sensitive to the contact angle and estimated a value of $\alpha=25^{\circ}$ as the most appropriate one for sandstone. Alternative testing configurations have also been proposed (see, e.g., \cite{Erarslan2012,Yu2021} and Refs. therein). The aforementioned studies provide material-specific estimations of test geometry (contact angles) that result in a stress state where the maximum tensile stress is attained at the centre of the disk. This can be achieved by using a sufficiently large jaw radius (sufficiently small contact angle). However, small contact angles result in high contact stresses that cause premature cracking near the loading region \cite{Komurlu2015}. Thus, finding a suitable testing configuration involves striking a balance between ensuring that the contact angle is both: (i) small enough such that the maximum tensile stress is attained at the centre and Eq. (\ref{eq:S1withLoad}) is satisfied, and (ii) large enough such that cracking does not occur in the compressive region beneath the jaw. This is not straightforward as it depends on a number of testing and material parameters and even today technical standards differ in their recommendations (see, e.g., Refs. \cite{ASTMD3697,Bieniawski1978}). There is a need for a generalised approach that will enable mapping the regimes of validity of the Brazilian test for arbitrary choices of material and test configuration.\\

In this work, we use the generalised Griffith criterion \cite{Griffith1924,Fairhurst1964} to gain insight into the location of crack initiation in the Brazilian test. By considering the entire failure envelope, we ensure that not only is the maximum tensile stress attained at the centre of the sample at the moment of failure but also that this crack nucleation event is not preceded by cracking elsewhere in the sample. Finite element calculations are conducted to build maps that enable assessing the experiment viability for any material and test geometry. First, we analyse the stress state at the disk centre as a function of the load and quantify the error associated with Hondros's solutions, Eqs. (\ref{eq:Hondros2}) and (\ref{eq:S1withLoad}), for relevant material properties and testing configurations. Second, we map the conditions that lead to crack nucleation at the disk centre and thus to a valid test. Calculations span the main classes of rocks and assess the suitability of current testing standards. We find that the range of conditions where a Brazilian test is valid is much narrower than previously thought. A protocol is presented to ensure that the experiment leads to a valid estimate of the material tensile strength. This is exemplified with specific case studies and facilitated by providing a MATLAB App that takes as input the test data and provides as output the validity of the experiment and the magnitude of the tensile strength.

\section{Generalised Griffith criterion for crack initiation}
\label{Sec:GriffithCriterion}

\begin{figure}[H]
    \centering
    \includegraphics[width=0.45\textwidth]{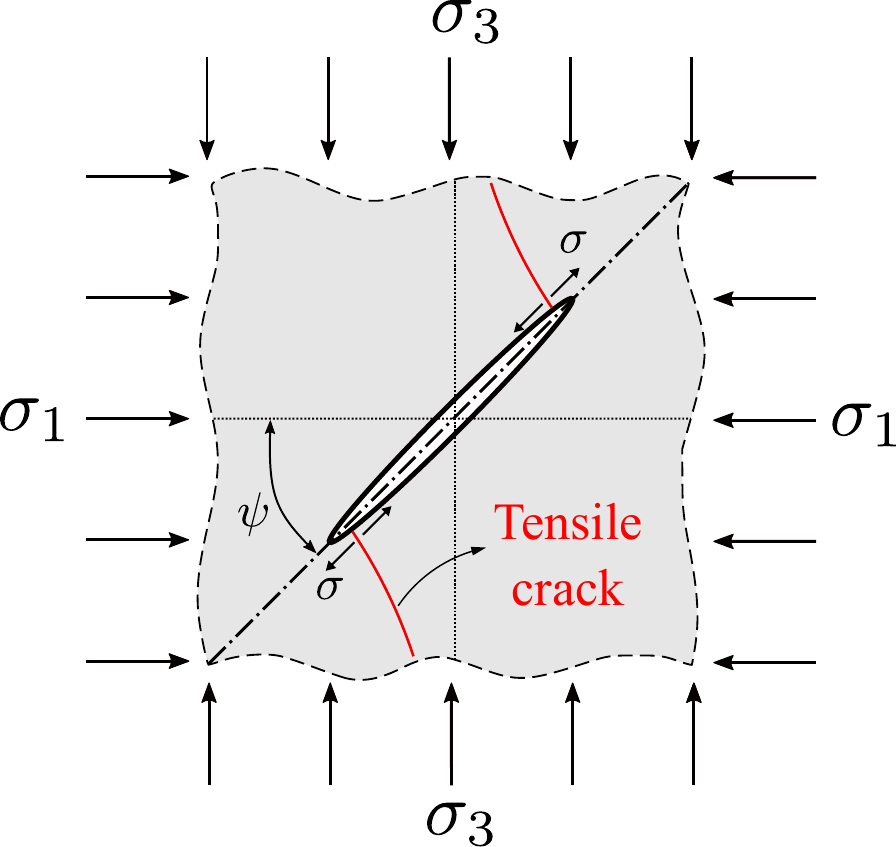}
    \caption{Local stress state in a Griffith micro-crack, with $\psi$ denoting the crack inclination angle. When the local tensile stresses reach the material tensile strength $\sigma_t$, wing cracks nucleate near the edges of the original micro-crack.}
    \label{Fig:InclinedCrack}
\end{figure}

Griffith \cite{Griffith1924} studied the fracture of brittle materials under compressive loads by assuming that the rupture process was driven by local flaws within the material. As shown in Fig. \ref{Fig:InclinedCrack}, local tensile stresses will develop near existing flaws when these are oriented at an angle relative to the principal directions of the applied stress. Denoting the major and minor principal stresses as $\sigma_1$ and $\sigma_3$, respectively, Griffith's \cite{Griffith1924} two-part criterion for the onset of fracture is given as follows,\footnote{A detailed derivation can be found in Ref. \cite{Jaeger2009}.}
\begin{equation}
\begin{cases}
\sigma_{1} =\sigma_t  \quad  & \text{if} \quad 3 \sigma_{1}+\sigma_{3} \geq 0 \\
(\sigma_{1}-\sigma_{3})^2 = -8 \sigma_t (\sigma_{1}+\sigma_{3})  \quad & \text{if} \quad  3 \sigma_{1}+\sigma_{3} < 0
\end{cases}
\label{Eq:Griffith criterion}
\end{equation}

\noindent with the initial crack orientation being respectively given by the angles: 
\begin{equation}
\begin{cases}
\psi =\pi/2  \quad  & \text{if} \quad 3 \sigma_{1}+\sigma_{3} \geq 0 \\
\psi  = \frac{1}{2} \cos^{-1} \left( \frac{\sigma_1 - \sigma_3}{2 (\sigma_1 + \sigma_3)} \right)   \quad & \text{if} \quad  3 \sigma_{1}+\sigma_{3} < 0
\end{cases}
\label{Eq:GriffithAngles}
\end{equation}

\begin{figure}[H]
    \centering
    \begin{subfigure}{0.6\textwidth}
    \includegraphics[width=\textwidth]{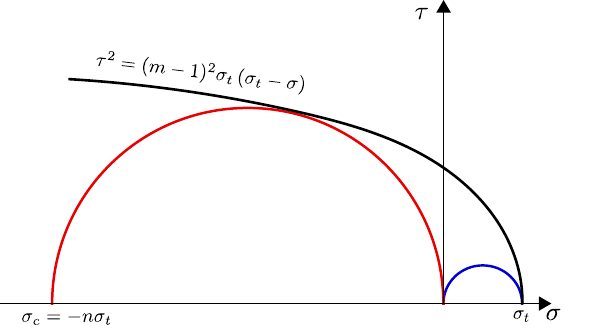}
    \caption{}
    \label{}
    \end{subfigure}
    \begin{subfigure}{0.6\textwidth}
    \includegraphics[width=\textwidth]{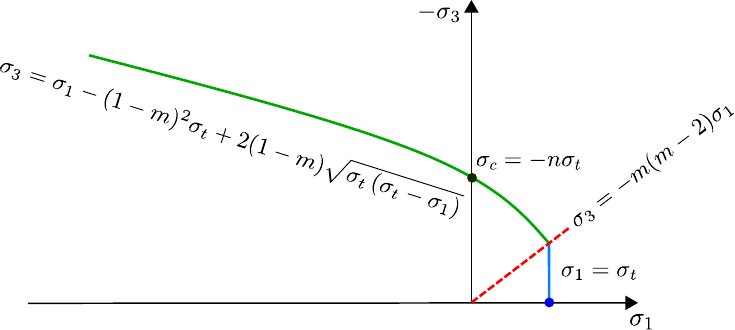}
    \caption{}
    \label{}
    \end{subfigure}
    \caption{Generalised Griffith criterion. Mohr diagram showing the generalised parabolic failure envelope in terms of: (a) normal $\sigma$ and shear $\tau$ stresses compression, and (b) minor $\sigma_3$ and major $\sigma_1$ principal stresses. Here, $n=-\sigma_c/\sigma_t$ and $m=\sqrt{n+1}$.}
    \label{fig:G-Griffith-1}
\end{figure}

Two aspects must be emphasised. First, it is observed that for a regime where $\sigma_1$ is tensile and $\sigma_3$ is compressive with an absolute value lower than three times $\sigma_1$, conditions of purely tensile failure take place, with cracks parallel to the original flaw \cite{Hoek2014}. Second, the criterion indicates that the material compressive strength $\sigma_c$ is eight times its tensile strength as Eq. (\ref{Eq:Griffith criterion})b gives $\sigma_3=\sigma_c=-8\sigma_t$ under uniaxial compression ($\sigma_1=0$). While this is of the right order of magnitude, it limits the application of the criterion to materials with a compressive-to-tensile strength ratio of 8. To overcome this and generalise Griffith's criterion, Fairhurst \cite{Fairhurst1964} proposed an extension to allow for arbitrary compression-to-tensile strength ratios. This is achieved by defining a parabolic Mohr envelope that encloses the uniaxial tensile and compressive strength circles, with the former being touched at its vertex and the latter being tangent to the envelope - see Fig. \ref{fig:G-Griffith-1}a. Accordingly, defining $n$ as the compressive-to-tensile strength ratio ($n=-\sigma_c/\sigma_t$), the relation describing the compressive strength circle is given by,
\begin{equation}
\left( \sigma+\frac{n \sigma_t}{2} \right)^2+\tau^2=\left( \frac{n \sigma_t}{2} \right)^2
\end{equation}

\noindent with $\sigma$ and $\tau$ respectively denoting the normal and shear stresses. 

In terms of the principal stress space, the generalised Griffith criterion reads:
\begin{equation}
\begin{cases}
\sigma_{1} =\sigma_t  \quad  & \text{if} \quad m(m-2) \sigma_{1}+\sigma_{3} \geq 0 \\
\sigma_3=\sigma_1-(1-m)^2\sigma_t+2(1-m)\sqrt{\sigma_t\left(\sigma_t-\sigma_1\right)}  \quad & \text{if} \quad  m(m-2) \sigma_{1}+\sigma_{3}  <  0 
\end{cases}
\label{Eq:FairhurstCriterion}
\end{equation}

\noindent where $m$ is a material parameter defined as $m=\sqrt{n+1}$. The failure envelope is shown graphically in Fig. \ref{fig:G-Griffith-1}b. The generalised Griffith criterion particularises to the original Griffith criterion (\ref{Eq:Griffith criterion}) for $n=8$ and otherwise extends it to arbitrary tensile and compressive material strengths. It is worth noting that the adoption of the generalised Griffith criterion necessarily implies that the Brazilian test is, generally, not a suitable experiment for measuring the tensile strength of materials with $n < 8$; see Eq. (\ref{Eq:FairhurstCriterion}a) and Fig. \ref{fig:G-Griffith-1}b and consider the fact that $\sigma_3 \approx -3\sigma_1$ at the disk centre for zero or small contact angles \cite{Jaeger2009}.

\section{The application of Griffith's criterion to the Brazilian test}
\label{Sec:GriffithBrazilian}

\begin{figure}[H]
    \centering
    \includegraphics[width=0.77\textwidth]{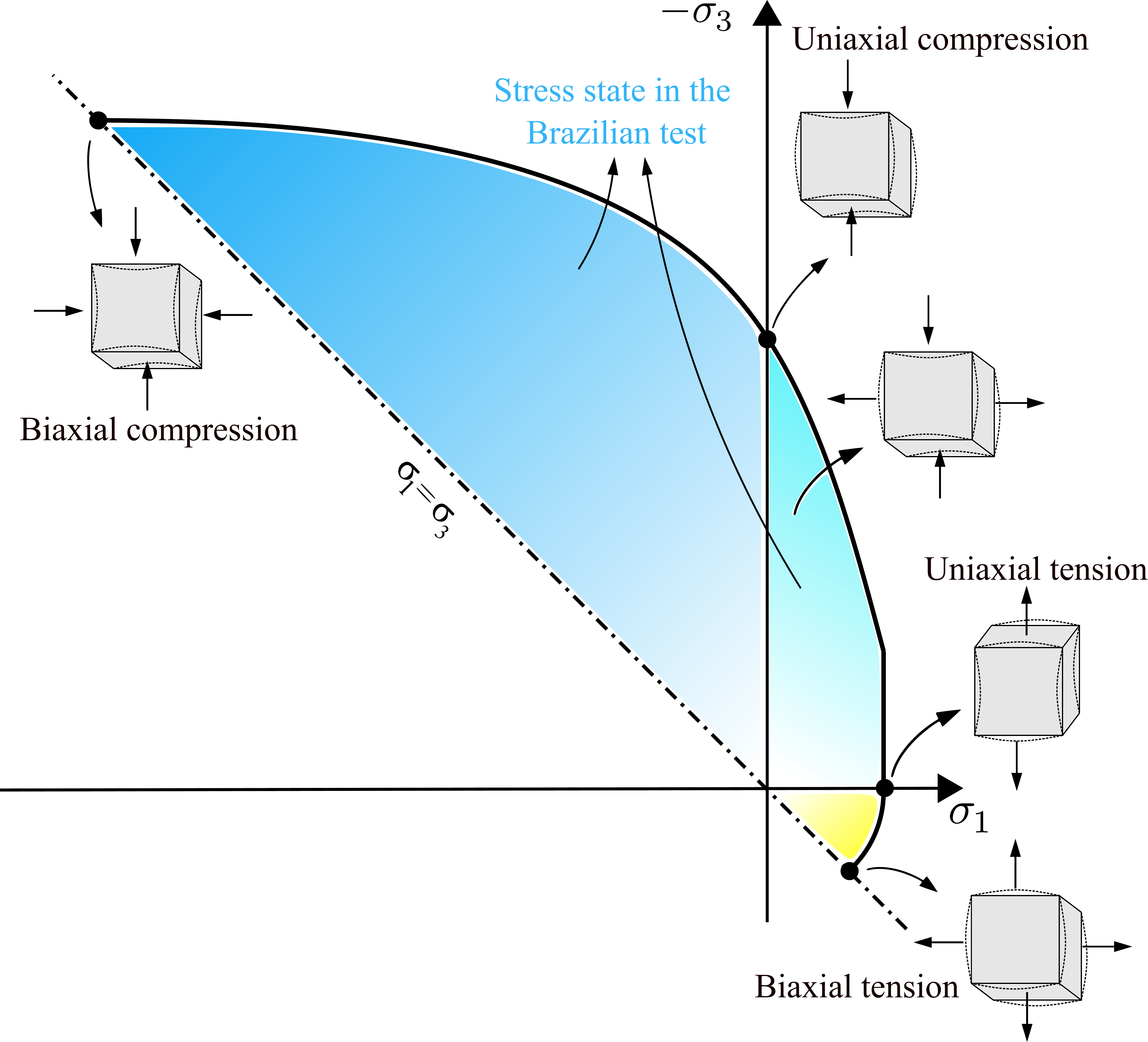}
    \caption{Stress states and typical failure envelope for rock-like materials, emphasising the two regimes relevant to the Brazilian test. The stress states are shown in the principle stress diagram, with tensile stresses being positive and $\sigma_1$ and $\sigma_3$ respectively denoting the major and minor principal stresses.}
    \label{fig:Stress-state-general for rock}
\end{figure}

During the Brazilian split test, the material points in the disk undergo a stress state that is characterised by two domains in the principal stress state - see Fig. \ref{fig:Stress-state-general for rock}. In some regions, such as in the vicinity of the jaws, material points exhibit compressive major and minor principal stresses ($\sigma_1<0$ \& $\sigma_3<0$). However, near the centre of the disk, the stress state is characterised by a maximum principal stress in tension ($\sigma_1>0$) and a minimum principal stress in compression ($\sigma_3<0$).

\begin{figure}[H]
    \centering
    \begin{subfigure}[t]{0.41\textwidth}
    \includegraphics[width=\textwidth]{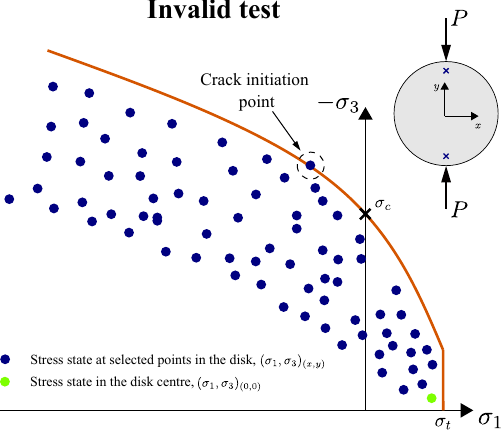}
    \caption{}
    \label{}
    \end{subfigure}\hspace{0.09\textwidth}
    \begin{subfigure}[t]{0.465\textwidth}
    \includegraphics[width=\textwidth]{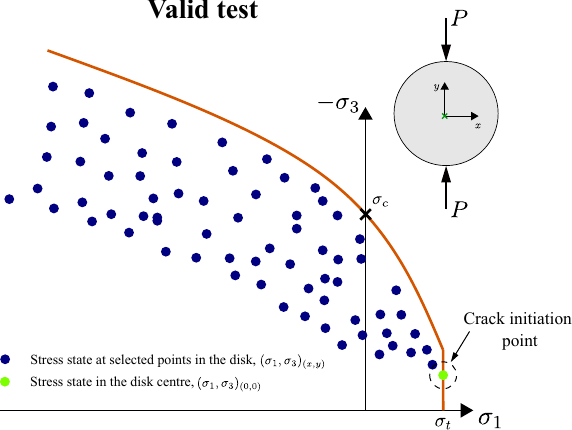}
    \caption{}
    \label{}
    \end{subfigure}
    \begin{subfigure}[t]{0.7\textwidth}
    \includegraphics[width=\textwidth]{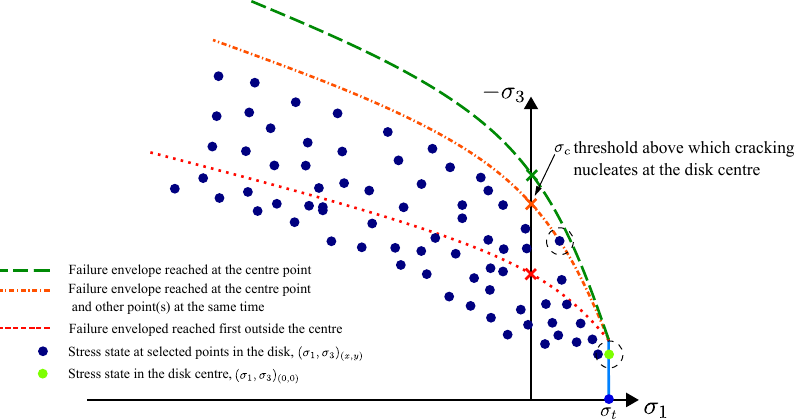}
    \caption{}
    \label{}
    \end{subfigure}
    \caption{Stress state at a discrete number of material points within the Brazilian disk and failure envelopes based on the generalised Griffith criterion. (a) Conditions leading to an invalid test; failure is attained outside from the disk centre. (b) Conditions leading to a valid test; $\sigma_1=\sigma_t$ at the centre of the sample ($0,0$). (c) Validity of the test as a function of the failure envelope ($\sigma_c$, $\sigma_t$) for a given stress state associated with a load $P$. A green dot is used to denote the stress state at the disk centre ($0,0$).}
    \label{fig:Stress-state-Graph}
\end{figure}

As discussed in Section \ref{Sec:Introduction}, the controversy surrounding the Brazilian test is related to the crack initiation location. For the experiment to provide a valid estimate of the material tensile strength, the onset of cracking must take place at the centre of the disk and the relation between the critical load and $\sigma_1$ at the disk centre must be known. One can use the failure envelope of the generalised Griffith criterion (Fig. \ref{fig:G-Griffith-1}) to analyse the stress state in the disk and map the conditions of validity. This is shown in a schematic manner in Fig. \ref{fig:Stress-state-Graph}, where a cloud of points is used to represent the potential stress states in a discrete number of material points distributed within the disk, $(\sigma_1, \sigma_3)_{(x,y)}$. Two scenarios can essentially occur. On the one hand, Fig. \ref{fig:Stress-state-Graph}a, the test is invalid if the first material point reaching the failure envelope is not located in the centre of the disk. This is, for example, what happens when cracking is observed close to the loading jaws. On the other hand, Fig. \ref{fig:Stress-state-Graph}b, if the failure envelope is reached first by the material point located at the disk centre ($x=0$, $y=0$), then a valid estimate of the tensile strength is obtained: $\sigma_t=(\sigma_1)_{(0,0)}$.\\

For a given applied load, test geometry and elastic properties of jaws and disk, the validity of the test will be determined by the failure envelope (i.e., the magnitude of $\sigma_c$ and $\sigma_t$). Fig. \ref{fig:Stress-state-Graph}c shows a scenario where one of the conditions of validity of the Brazilian test has been met: the centre of the disk (green dot) is in a stress state where $(\sigma_1)_{(0,0)}=\sigma_t$. However, the test is still not valid if the ratio $\sigma_c/\sigma_t$ is sufficiently low - several material points are above the envelope, implying that failure has occurred elsewhere at a smaller load. This scenario is illustrated with a red dotted curve in Fig. \ref{fig:Stress-state-Graph}c. If the ratio $\sigma_c/\sigma_t$ is sufficiently large (green dashed curve), then the only point in contact with the envelope is the centre one, and the experiment is valid.

The limiting case is that where the failure envelope is met at two or more points at the same time, one of which is located at the disk centre. This is illustrated in Fig. \ref{fig:Stress-state-Graph}c with an orange dash-dotted line and provides the threshold of admissible $\sigma_c/\sigma_t$ ratios for a Brazilian test to be valid. Thus, for a given load, geometry and material parameters, one can use numerical analysis to estimate the stress state at any point in the disk $(\sigma_1,\sigma_3)_{(x,y)}$ and utilise the generalised Griffith criterion to determine the compressive strength associated with a failure envelope passing through that point; i.e., re-arranging Eq. (\ref{Eq:FairhurstCriterion})b:
\begin{equation}
(\sigma_{c})_{(x,y)}=-\mathrm{\sigma_t}\,\left(\frac{{\left(\mathrm{\sigma_t}-\sqrt{\mathrm{\sigma_t}\,\left(\mathrm{\sigma_t}-(\sigma_{1})_{(x,y)}\right)}+\sqrt{\mathrm{\sigma_t}\,\left(\mathrm{\sigma_t}-(\sigma_{3})_{(x,y)}\right)}\right)}^2}{{\mathrm{\sigma_t}}^2}-1\right)
\label{Eq:G-Com}
\end{equation}

For the failure condition to be first met at the disk centre, the maximum value of $(\sigma_{c})_{(x,y)}$ among all material points in the disk, as estimated \textit{via} Eq. (\ref{Eq:G-Com}), must be equal or smaller than the real material compressive strength $\sigma_c$. Hence, since $\sigma_c$ is a known material property that can be measured independently, one can combine numerical analysis and the generalised Griffith's criterion to map the conditions that lead to failure initiation from the centre of the disk. In this way, the two validity conditions of the Brazilian test - cracking initiating at the centre ($0,0$) and $(\sigma_1)_{(0,0)}=\sigma_t$ - can be incorporated in the analysis, as shown below.

\section{Analysis}
\label{Sec:Results}

We proceed to combine finite element analysis and the generalised Griffith criterion to map the regimes of validity of the Brazilian test.

\subsection{Preliminaries}

The location of crack initiation in the Brazilian test is a function of 2 geometrical and 6 material parameters: the jaw radius ($R_j$), the disk radius ($R_d$), the elastic properties of the disk ($E_d$, $\nu_d$) and jaws ($E_j$, $\nu_j$), and the tensile ($\sigma_t$) and compressive ($\sigma_c$) strengths of the material being tested. Assuming that cracking initiates along the vertical middle axis of the disk, the crack initiation location can be fully characterised by a variable $Y$, equal to 0 at the centre and to $R_d$ at the edge. Then, dimensional analysis dictates that the solution is a function of the following non-dimensional sets:
\begin{equation}
\frac{Y}{R_d}=F \left( \frac{R_j}{R_d},\frac{E_j}{E_d},\nu_j,\nu_d,\frac{\sigma_c}{E_d},\frac{\sigma_t}{E_d} \right) \, .
\label{eq:crack-location-dim}
\end{equation}

Further assuming that crack nucleation takes place at the centre of the disk ($Y/R_d=0$), as required for the test to be valid, then Eq. (\ref{eq:crack-location-dim}) can be re-arranged to:
\begin{equation}\label{eq:crack-location-dim2}
\frac{\sigma_c}{\sigma_t}=G \left( \frac{R_j}{R_d},\frac{E_j}{E_d},\nu_j,\nu_d,\frac{\sigma_c}{E_d} \right) \, .
\end{equation}

\noindent Thus, conducting calculations over relevant ranges of the five non-dimensional sets in Eq. (\ref{eq:crack-location-dim2}) will enable mapping the conditions that lead to cracking at the disk centre.\\  

\begin{figure}[H]
    \centering
    \begin{subfigure}[htp]{0.45\textwidth}
    \centering
    \includegraphics[width=\textwidth]{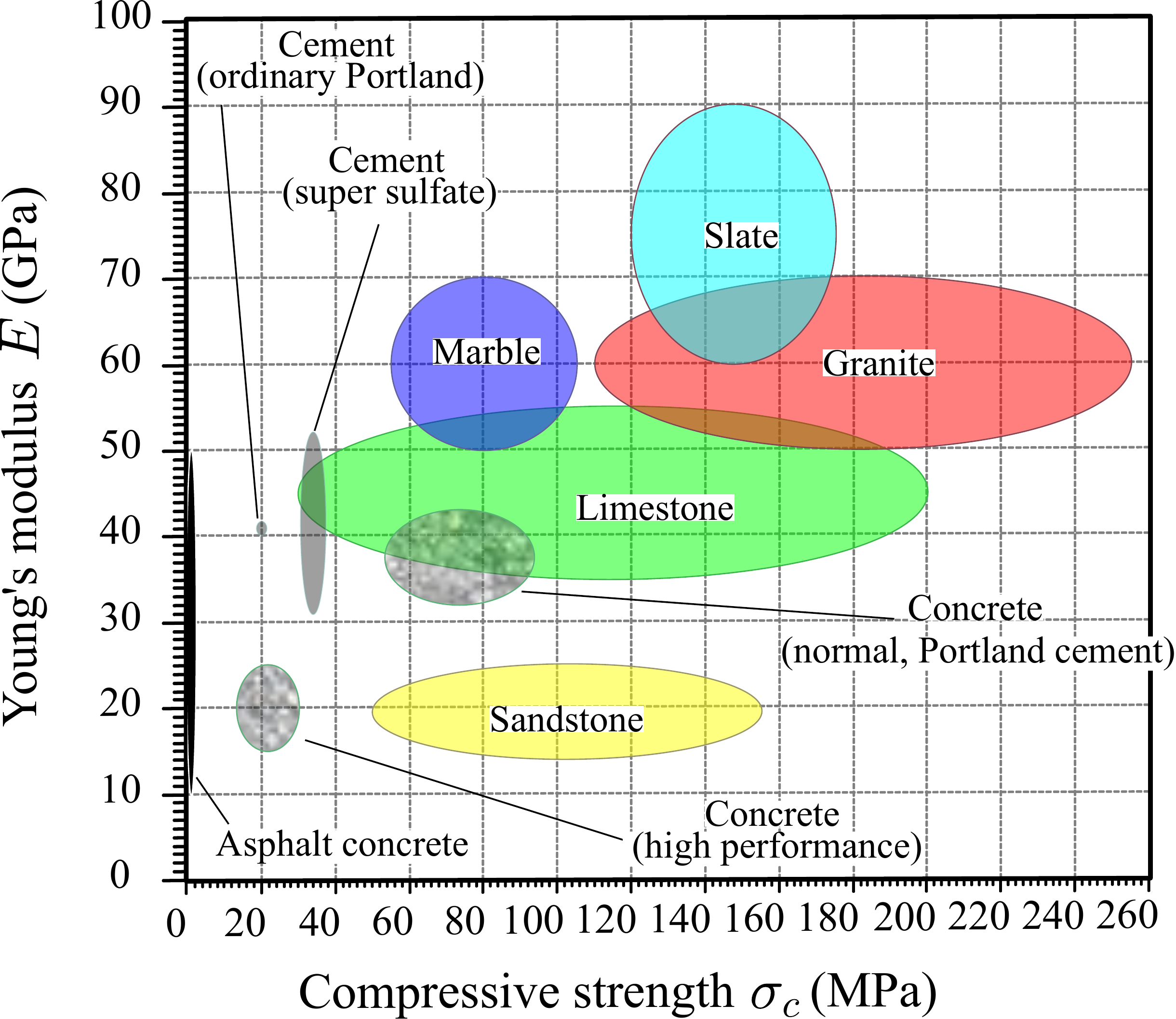}
    \caption{}
    \label{}
    \end{subfigure}\hspace{.05 \textwidth}
    \begin{subfigure}[htp]{0.45\textwidth}
    \centering
    \includegraphics[width=\textwidth]{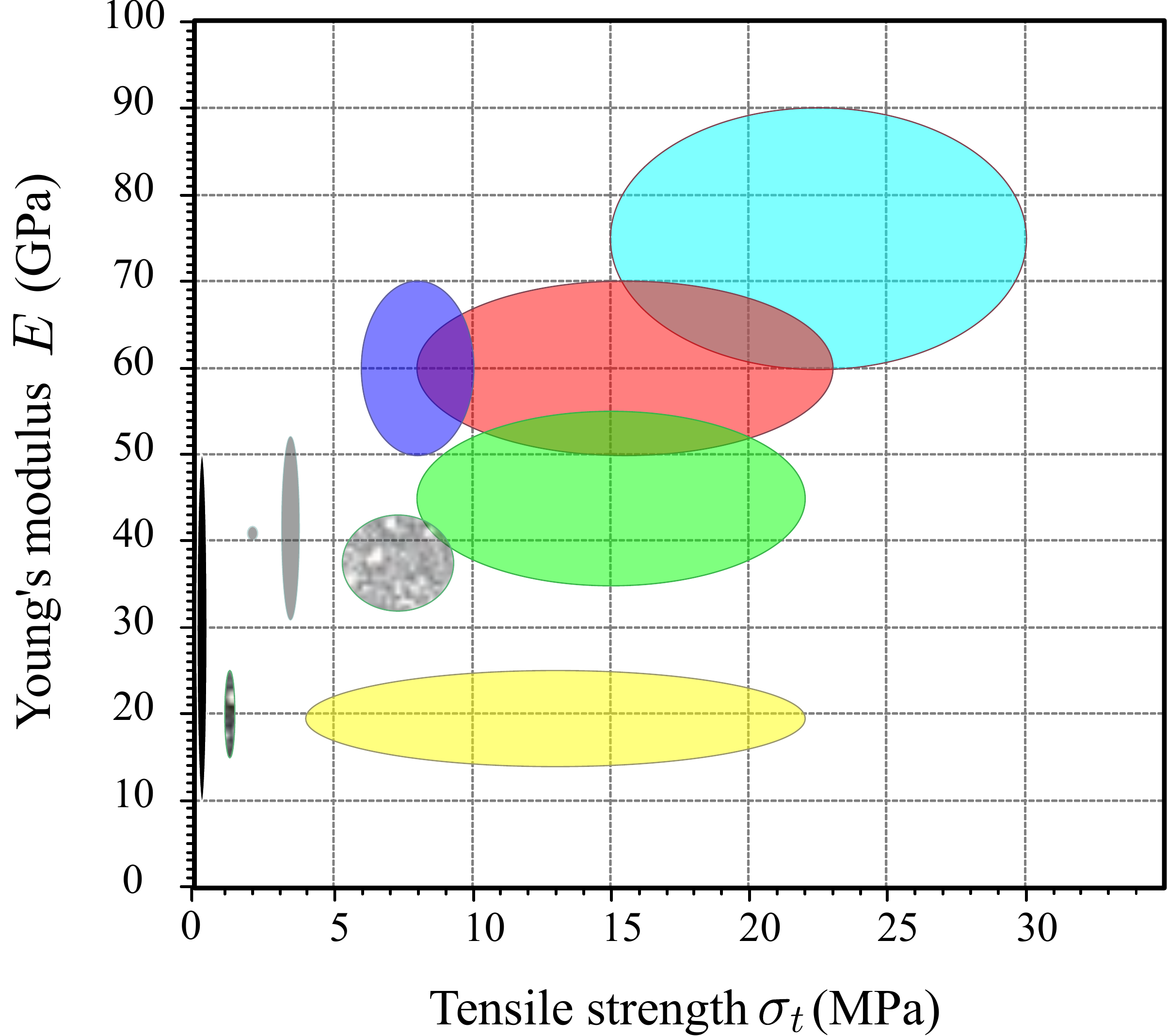}
    \caption{}
    \label{}
    \end{subfigure}
    \begin{subfigure}[htp]{0.45\textwidth}
    \centering
    \includegraphics[width=\textwidth]{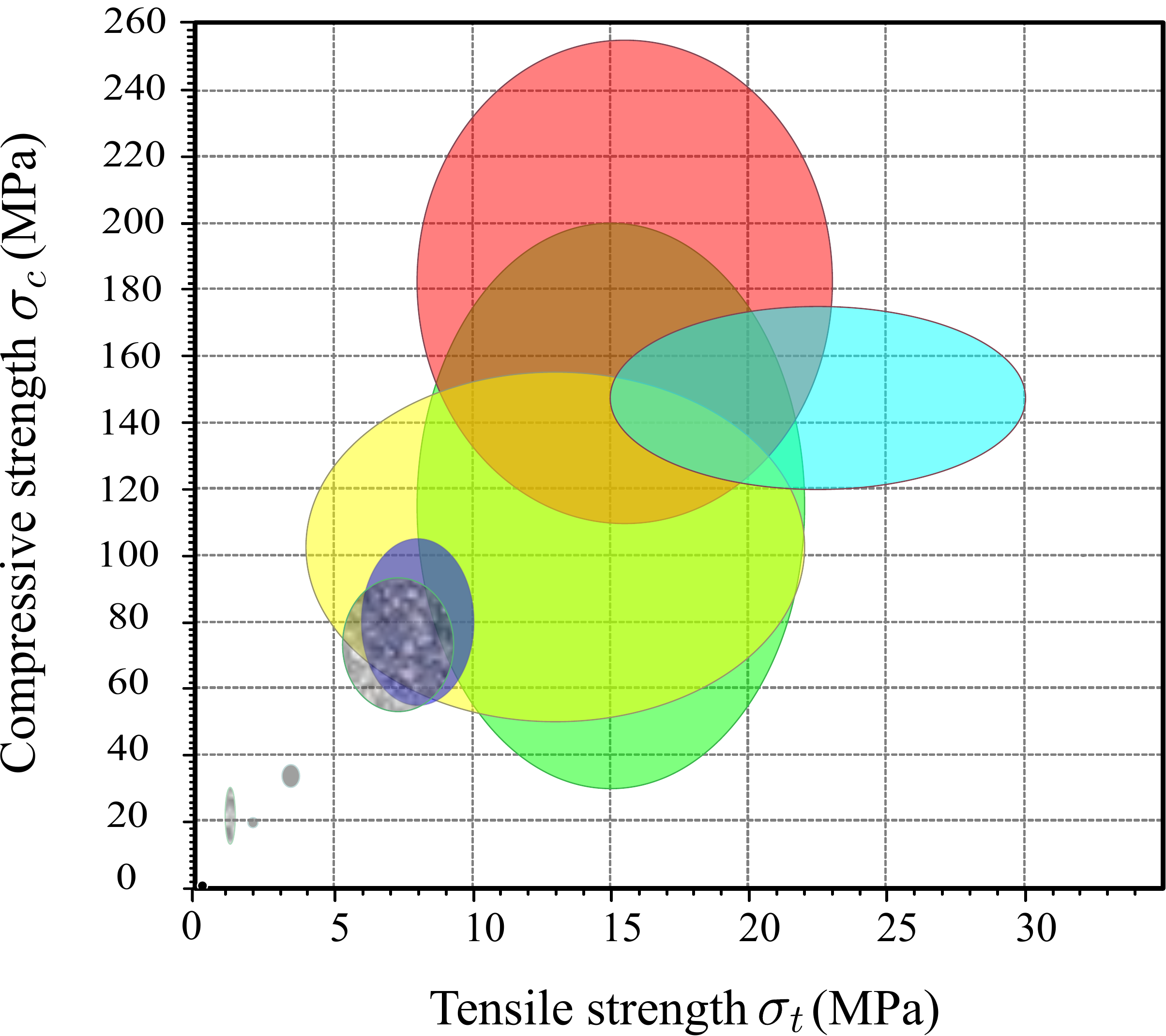}
    \caption{}
    \label{}
    \end{subfigure}\hspace{.05 \textwidth}
    \begin{subfigure}[htp]{0.46\textwidth}
    \centering
   \includegraphics[width=\textwidth]{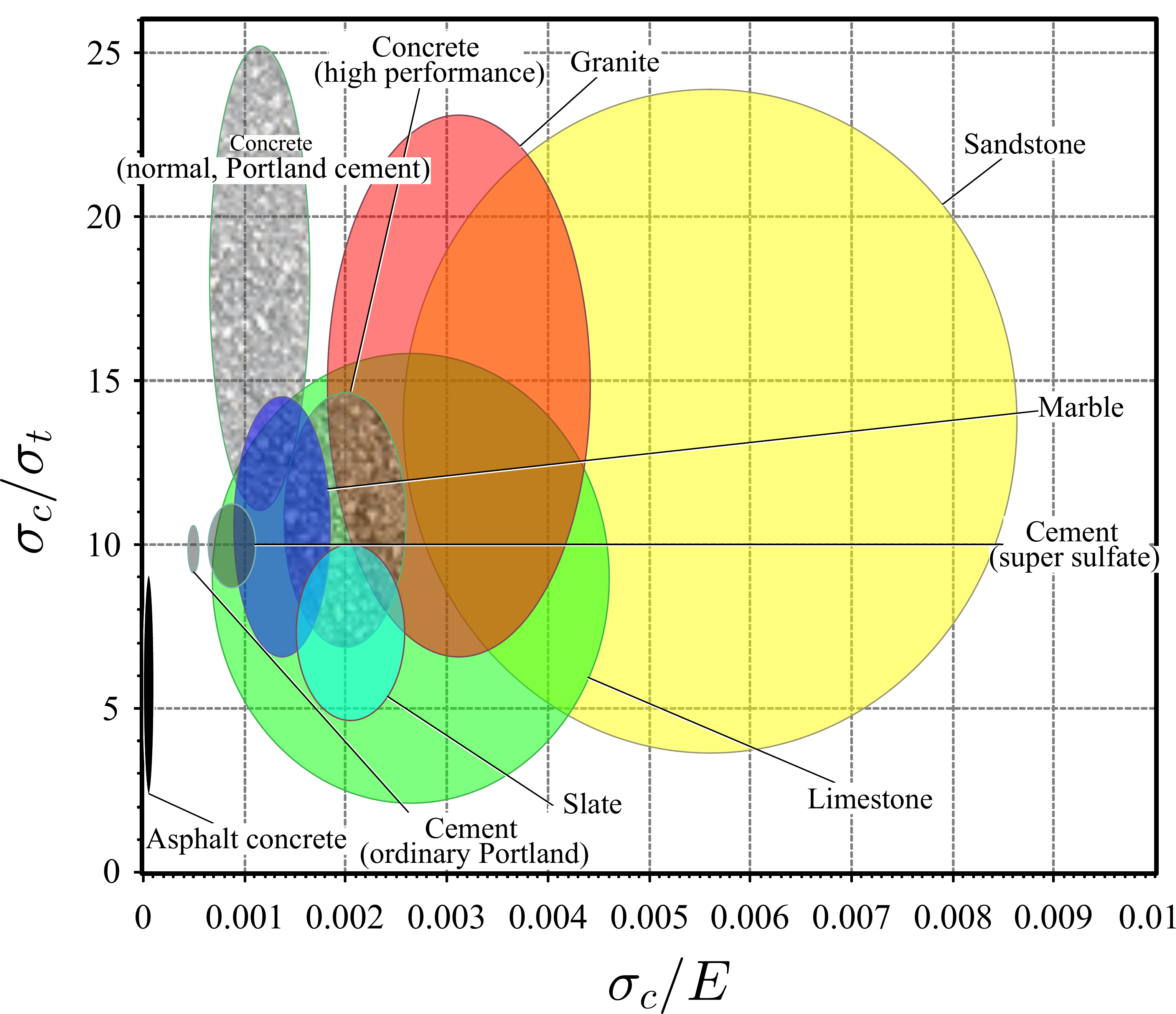}
   \caption{}
    \label{fig:fcE-fcft-total}
    \end{subfigure}
    \caption{Material property range of rock-like materials. Ashby charts showing the relations between (a) Young's modulus ($E$) and compressive strength ($\sigma_c$), (b) Young's modulus ($E$) and tensile strength ($\sigma_t$), (c) compressive ($\sigma_c$) and tensile ($\sigma_t$) strengths, and (d) ratio of compressive-to-tensile strength $(\sigma_c/\sigma_t)$ and ratio of compressive strength to elasticity modulus $\sigma_c/E$. The data is taken from the GRANTA Material library \cite{GRANTA2021} for granite, slate, marble, sandstone, limestone, concrete, cement and asphalt. The typical ranges for the Poisson's ratio of these materials are: granite $\nu=0.15-0.26$, slate $\nu=0.22-0.3$, marble $\nu=0.14-0.22$, sandstone $\nu=0.22-0.29$, limestone $\nu=0.2-0.26$, concrete $\nu=0.1-0.2$, cement $\nu=0.2-0.24$, and asphalt $\nu=0.35-0.36$.}
    \label{fig:ElasticProperties}
\end{figure}

We use the GRANTA Material library \cite{GRANTA2021} to define a suitable range of material properties. The Young's modulus, Poisson's ratio, tensile strength and compressive strength of the most widely used rock-like materials are shown in Figs. \ref{fig:ElasticProperties}a-\ref{fig:ElasticProperties}c. To conduct a comprehensive analysis, we vary the Young's modulus of the disk from 5 to 150 GPa. Also, Poisson's ratio is varied within the range 0.1 to 0.4. The jaws are typically made of steel and thus the following elastic properties are assumed: $E_j=210$ GPa and $\nu=0.3$. Given that $E_j$ and $\nu_j$ are fixed (and known), the dimensional analysis conducted above suggests that the two critical non-dimensional sets are $\sigma_c/\sigma_t$ and $\sigma_c/E_d$. Thus, we proceed to plot their relationship for a wide range of materials in Fig. \ref{fig:ElasticProperties}d. It can be observed that relevant ranges of $\sigma_c/\sigma_t$ and $\sigma_c/E_d$ are approximately 2-30 and 0.0001-0.01, respectively.\\

To determine the stress state within the disk we conduct finite element analysis of the contact between the jaws and the sample and the subsequent material deformation. The commercial finite element package ABAQUS is used. Only one quarter of the test is simulated, taking advantage of symmetry. The radius of the disk equals $R_d=10$ mm while the jaw radius is varied from $R_j=11$ mm to the case of a flat jaw geometry ($R_j \to \infty$). Quadratic quadrilateral finite elements with full integration are used to discretise the disk and the jaw. Plane strain conditions are assumed. After a sensitivity analysis, a total of 28,241 elements are used to discretise the disk and between 4,102 and 4,459 elements are used for the jaw. The mesh is particularly fine in the disk and in the regions of the jaw that are in contact with the disk. A uniform negative vertical displacement is applied at the top of the jaw and the resulting reaction force is measured. The contact behaviour is modelled as follows. For the normal behaviour, we consider surface-to-surface hard contact, where Lagrangian multipliers are used to ensure that the contact pressure and the contact constraint minimise overclosure. For the tangential behaviour, frictionless contact is generally assumed although the role of friction is also investigated (see Section \ref{Sec:Effect of friction}), revealing a negligible influence.

\subsection{Mapping the stress state at the disk centre}
\label{Sec:AnalysisStressCentre}

\begin{figure}[H]
    \centering
    \begin{subfigure}{0.49\textwidth}
    \includegraphics[width=\textwidth]{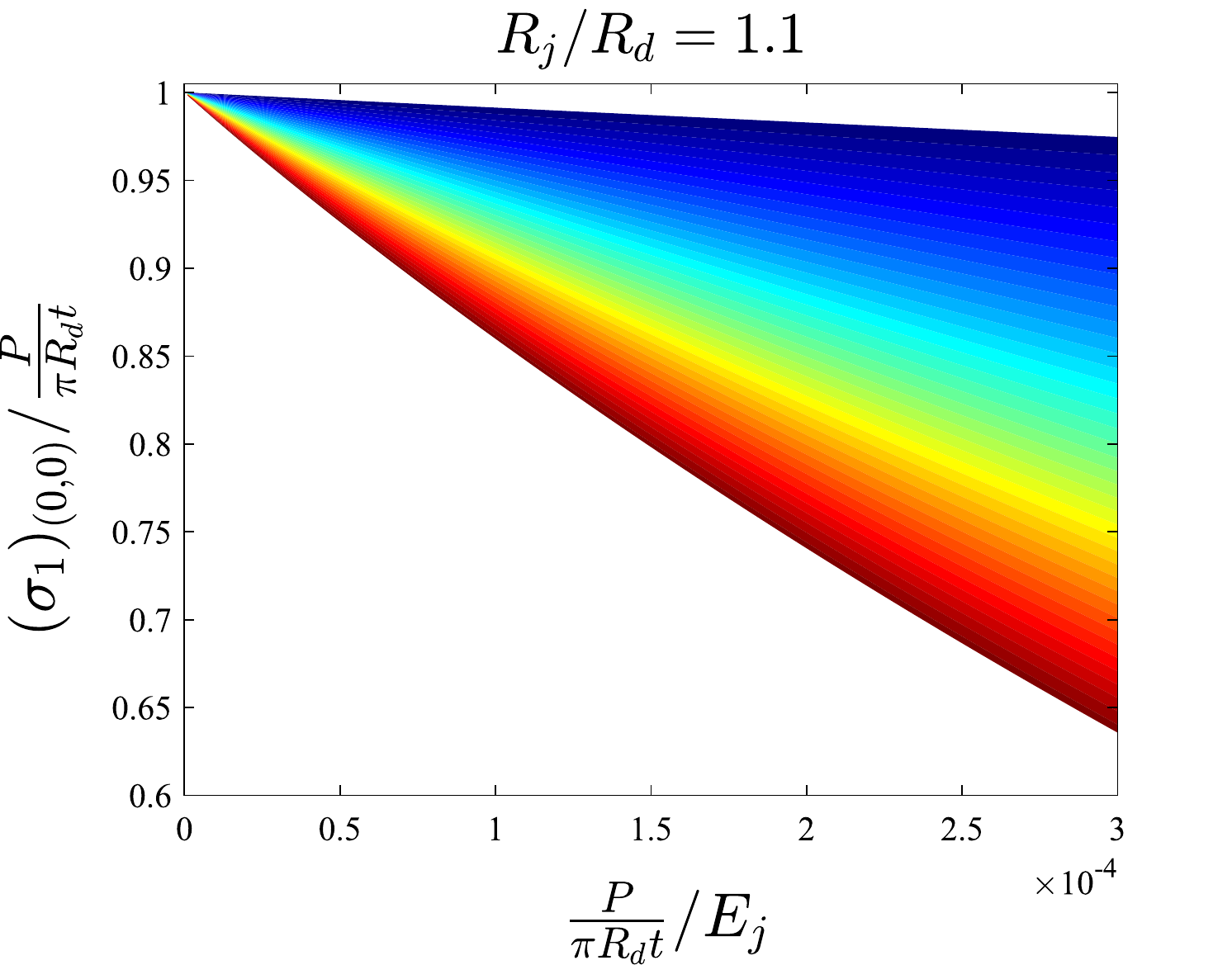}
    \caption{}
    \label{fig:load-R-Constant-a}
    \end{subfigure}
    \begin{subfigure}{0.49\textwidth}
    \includegraphics[width=\textwidth]{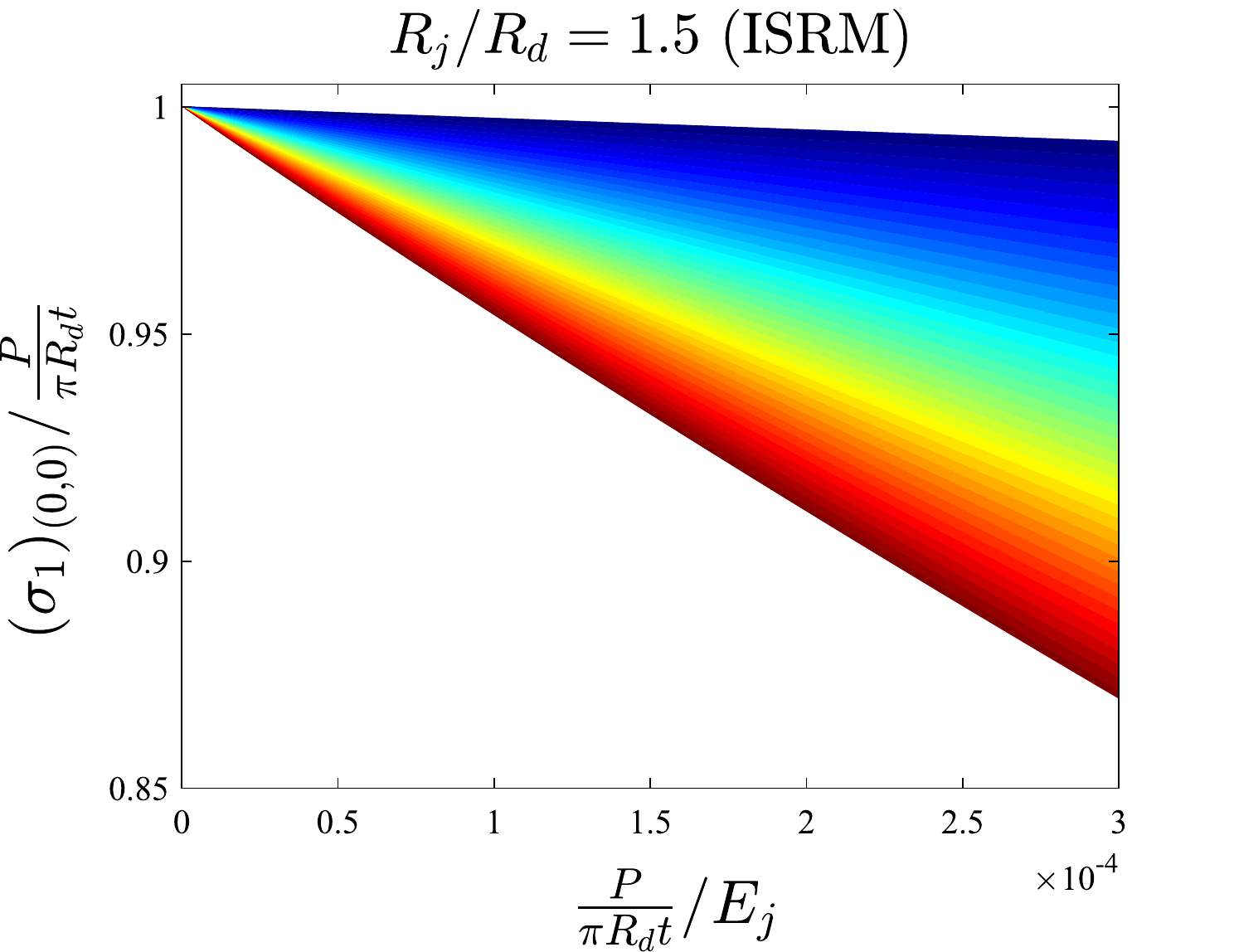}
    \caption{}
    \label{fig:load-R-Constant-b}
    \end{subfigure}\vspace{5 mm}
    \begin{subfigure}{0.57\textwidth}
    \includegraphics[width=\textwidth]{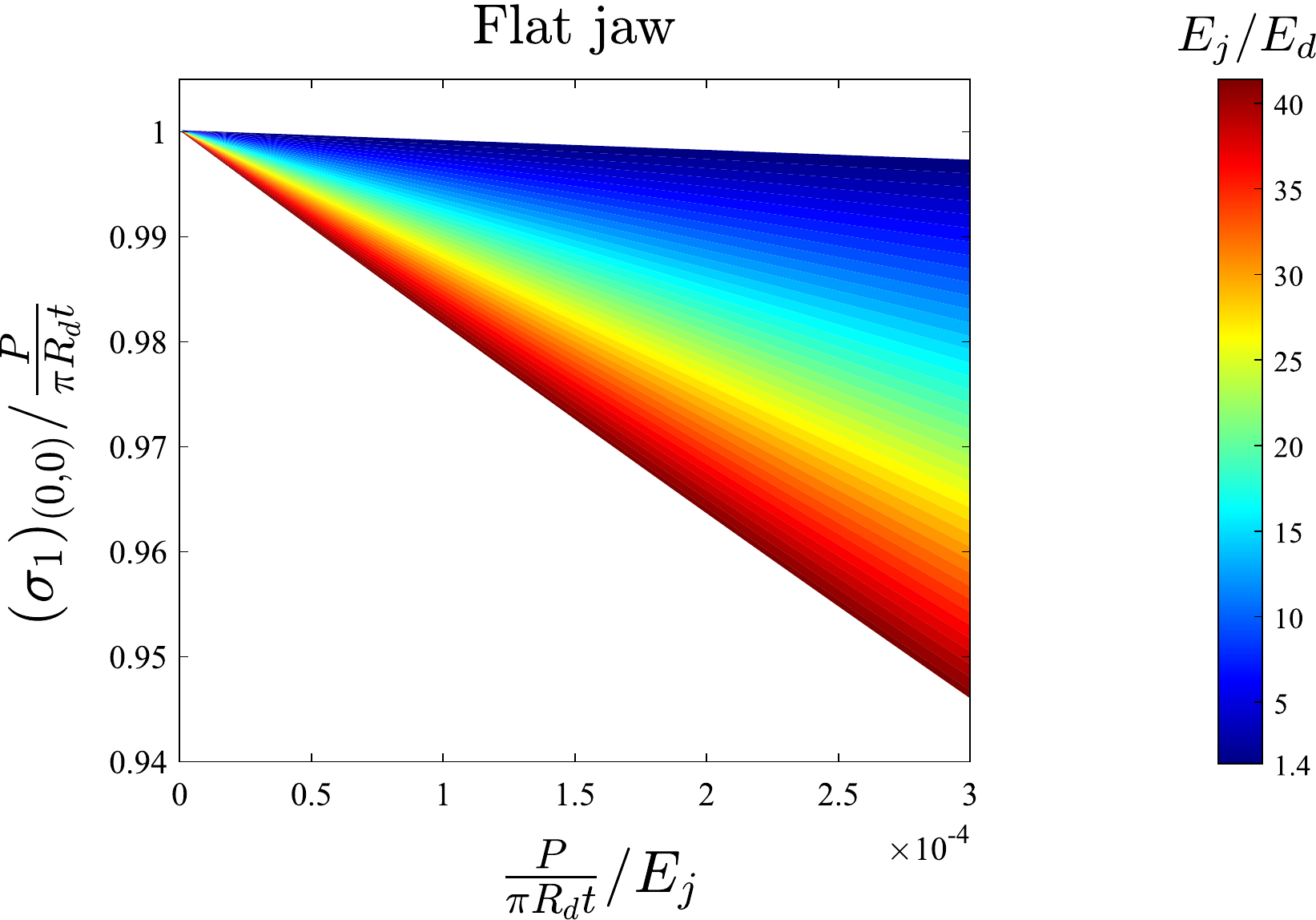}
    \caption{}
    \label{fig:load-R-Constant-c}
    \end{subfigure}
    \caption{Maps to quantify the stress state at the disk centre as a function of the material properties, test geometry and critical load. Normalised major principal stress versus dimensionless load for a wide range of $E_j/E_d$ values and the following test geometries: (a) $R_j/R_d=1.1$, (b) $R_j/R_d=1.5$ (as recommended by ISRM), and (c) flat jaws (as recommended by ASTM). Poisson's ratio in the disk is taken to be $\nu_d=0.2$.}
    \label{fig:load-R-Constant}
\end{figure}

We shall start by quantifying the relationship between the load $P$ and the stress state at the centre of the disk under a wide range of conditions. The goal is to map the scenarios where Eqs. (\ref{eq:S1withLoad}) and (\ref{eq:Hondros2}) are valid. We shall start by assessing the validity of Eq. (\ref{eq:S1withLoad}), an intrinsic assumption in the standards. The finite element results obtained are shown in Fig. \ref{fig:load-R-Constant} in terms of the stress state at the centre of the disk ($x=0,y=0$) versus the load for a wide range of $E_j/E_d$ values and selected choices of jaw radius, as given by the ratio $R_j/R_d$. In terms of test geometry, three scenarios are considered: $R_j/R_d=1.1$, $R_j/R_d=1.5$ (as in the ISRM standard) and flat jaws (one of the configurations recommended by the ASTM standard). The limits of the $x$-axis are chosen so as to encompass a wide range of realistic contact angles; the upper limit ($P/(\pi R_d t)=0.0003E_j$) corresponds to a tensile strength of roughly 60 MPa if a steel jaw ($E_j=210$ GPa) is considered in Eq. (\ref{eq:S1withLoad}), which is sufficiently high to cover the vast majority of rock-like materials.\\

The results reveal that Eq. (\ref{eq:S1withLoad}) is only valid for low load magnitudes and small $E_j/E_d$ ratios. The error is particularly significant for low $R_j/R_d$ values - note the $y$ axis limits in Fig. \ref{fig:load-R-Constant}a. But even for the case of flat jaws, as recommended by the ASTM standard, $(\sigma_1)_{(0,0)}/(P/\pi R_d t)$ is only equal to 1 for low contact angles (low $P$) and small Young's modulus mismatch. Consider for example a sandstone with $E_d=20$ GPa ($E_j/E_d=10.5$) and tensile strength $\sigma_t=20$ MPa ($P/(\pi R_d t) \approx 0.0001E_j$), see Fig. \ref{fig:ElasticProperties}; in all cases Eq. (\ref{eq:S1withLoad}) is not fulfilled, with the errors being of roughly 5\%, 2\% and 0.5\% for, respectively, the cases of $R_j/R_d=1.1$, $R_j/R_d=1.5$ (as suggested by ISRM) and flat jaws (as suggested by the ASTM standard). The maximum errors observed for these three configurations, relevant to materials with high tensile strength and low stiffness, are respectively 36\%, 13\% and 5\%. However, these maps enable a precise determination of the stress state in the centre of the disk and, accordingly, of the material tensile strength $\sigma_t$. One can use them to assess if the error intrinsic to the adoption of the point load equation is admissible, or directly as a replacement to Eq. (\ref{eq:S1withLoad}), as these maps enable determining the precise value of $\sigma_1(=\sigma_t)$ at the disk centre as a function of the material properties, test geometry and critical load.\\

\begin{figure}[H]
    \centering
    \begin{subfigure}[t]{0.43\textwidth}
    \includegraphics[width=\textwidth]{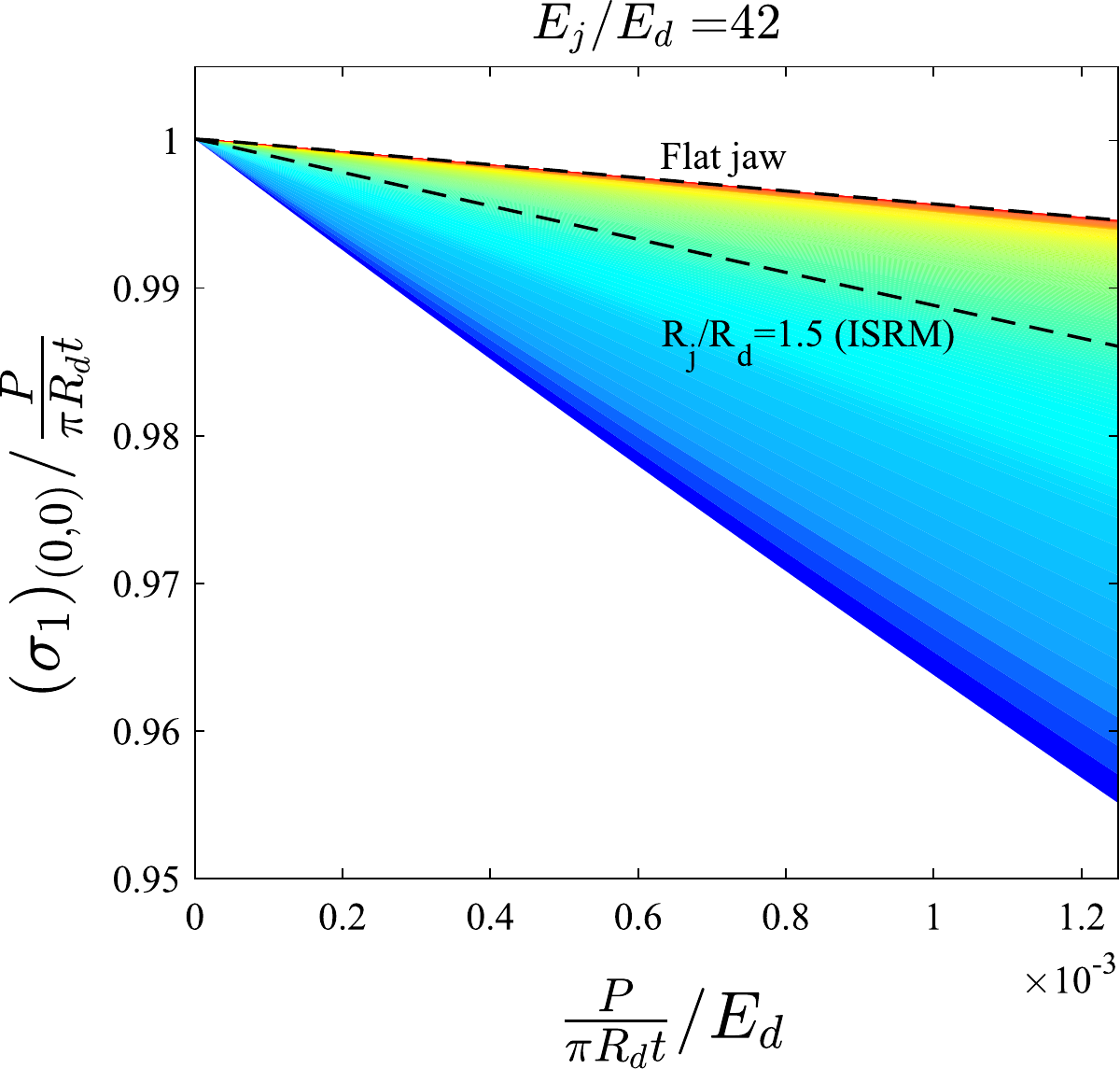}
    \caption{}
    \label{}
    \end{subfigure}\hspace{4 mm}
    \begin{subfigure}[t]{0.43\textwidth}
    \includegraphics[width=\textwidth]{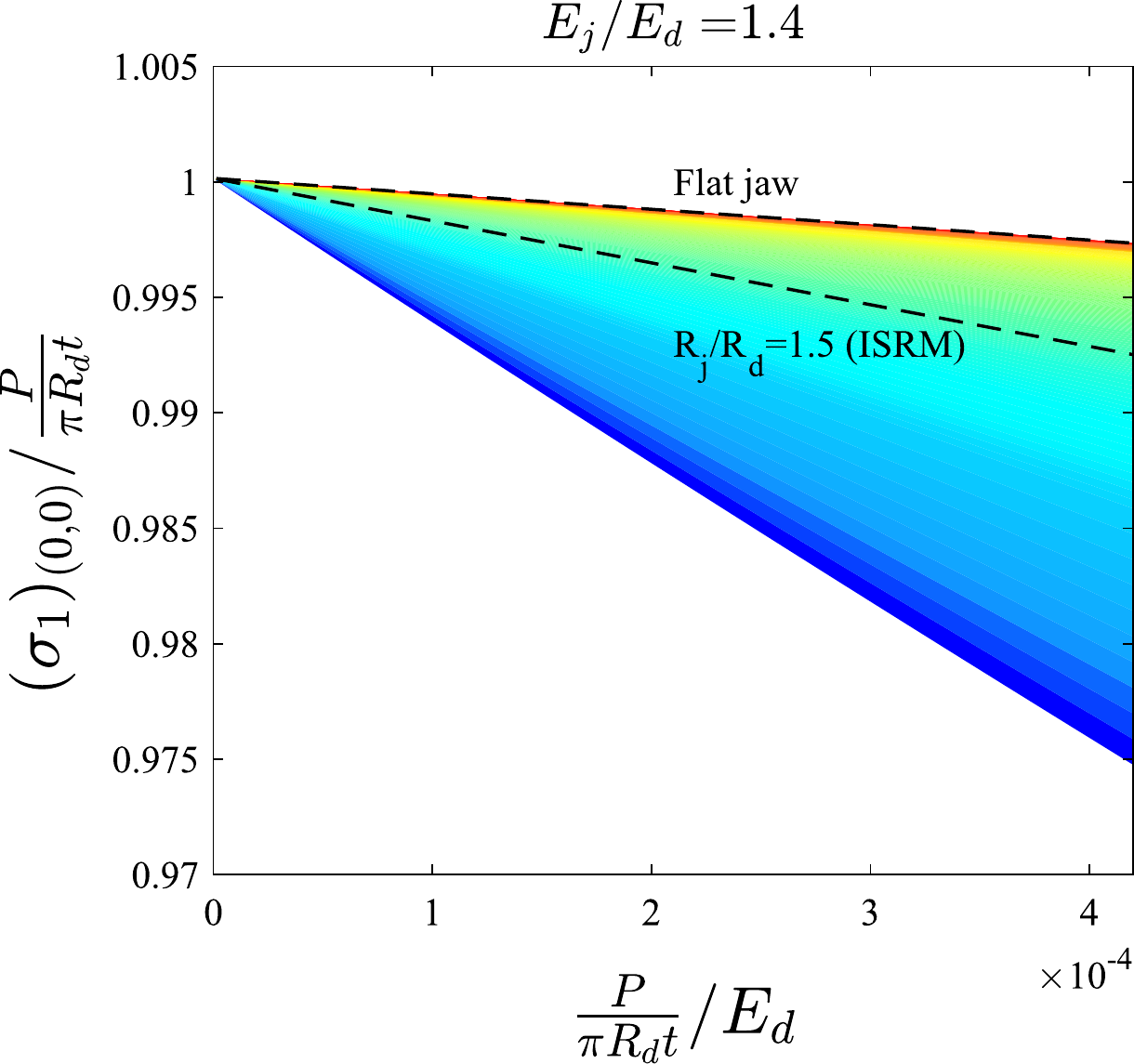}
    \caption{}
    \label{}
    \end{subfigure}\hspace{1 mm}
    \begin{subfigure}[t]{0.088\textwidth}\vspace{-73 mm}
    \includegraphics[width=\textwidth]{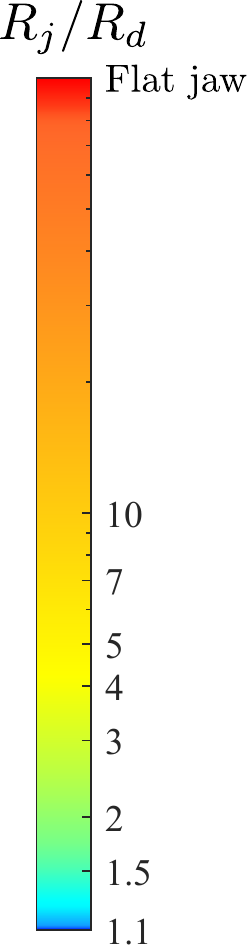}
    \end{subfigure}
    \caption{Maps to quantify the stress state at the disk centre as a function of the material properties, test geometry and critical load. Normalised major principal stress versus dimensionless load for a wide range of $R_j/R_d$ values and the following bounds of the elastic stiffness: (a) $E_j/E_d=42$, and (b) $E_j/E_d=1.4$. Poisson's ratio in the disk is taken to be $\nu_d=0.2$.}
    \label{fig:load-E-Constant}
\end{figure}

The results obtained for a wide range of jaw radii are given in Fig. \ref{fig:load-E-Constant}. Maps are provided as a function of the normalised load, using the Young's modulus of the rock as normalising parameter. Two figures are shown, corresponding to the lower and upper bounds of the elastic modulus; $E_j/E_d=42$ ($E_d \approx 5$ GPa, Fig. \ref{fig:load-E-Constant}a) and $E_j/E_d=1.4$ ($E_d \approx 150$ GPa, Fig. \ref{fig:load-E-Constant}b). Maps for other scenarios are provided in the Supplementary Material, so that experimentalists can accurately determine the stress state at the disk centre for arbitrary materials and test conditions. See also the Matlab App described in \ref{App:MATLABapp}. In agreement with expectations and with the results shown in Fig. \ref{fig:load-R-Constant}, stiffer materials bring the stress state close to that fulfilling Eq. (\ref{eq:S1withLoad}). Also, the error is relatively small when large jaw radii are used, with the limiting case being given by the flat jaws recommended by ASTM \cite{ASTMD3697}.

\begin{figure}[H]
    \centering
    \begin{subfigure}[t]{0.43\textwidth}
    \includegraphics[width=\textwidth]{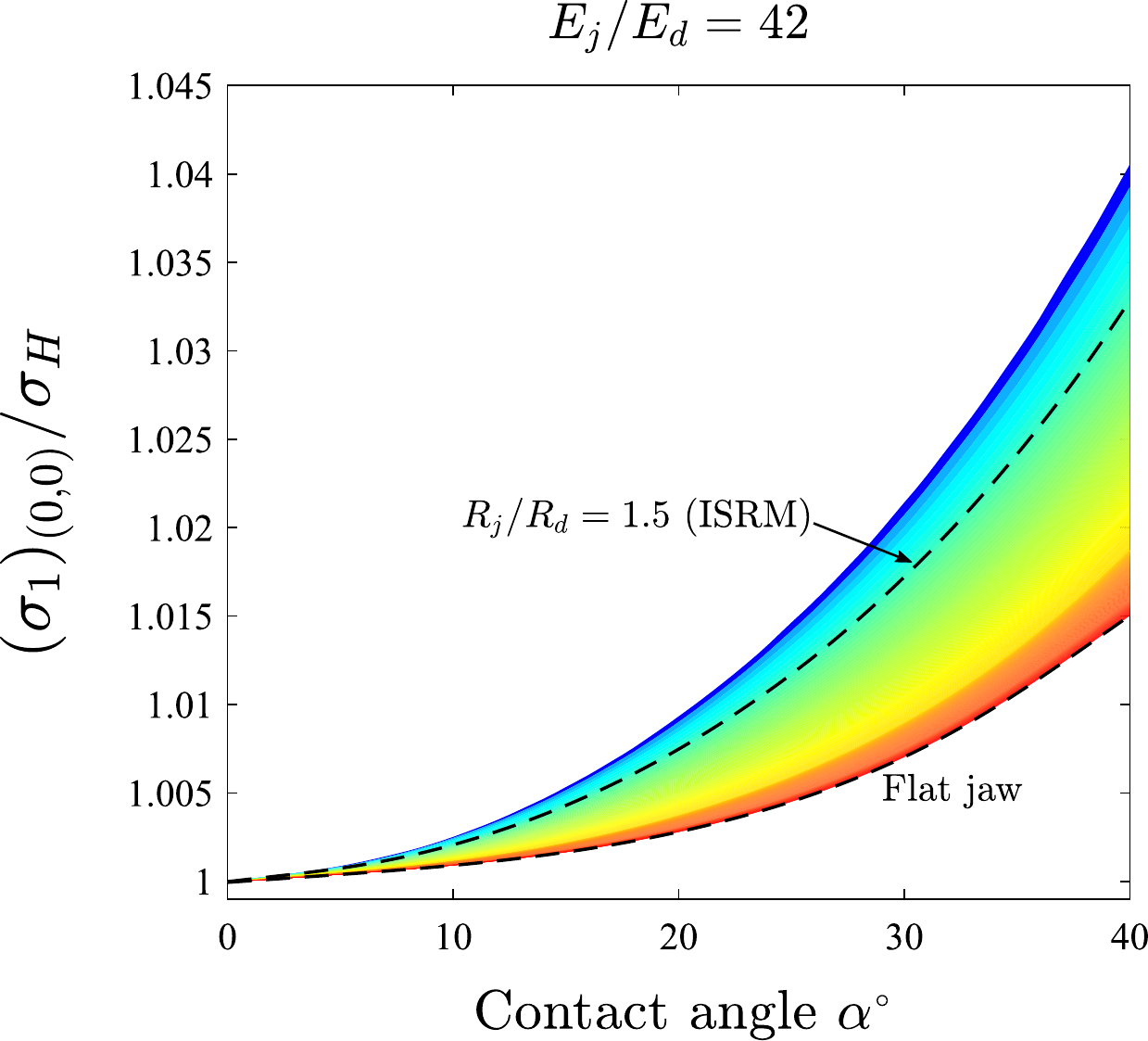}
    \caption{}
    \label{fig:H-E-Constant-a}
    \end{subfigure}\hspace{4 mm}
    \begin{subfigure}[t]{0.43\textwidth}
    \includegraphics[width=\textwidth]{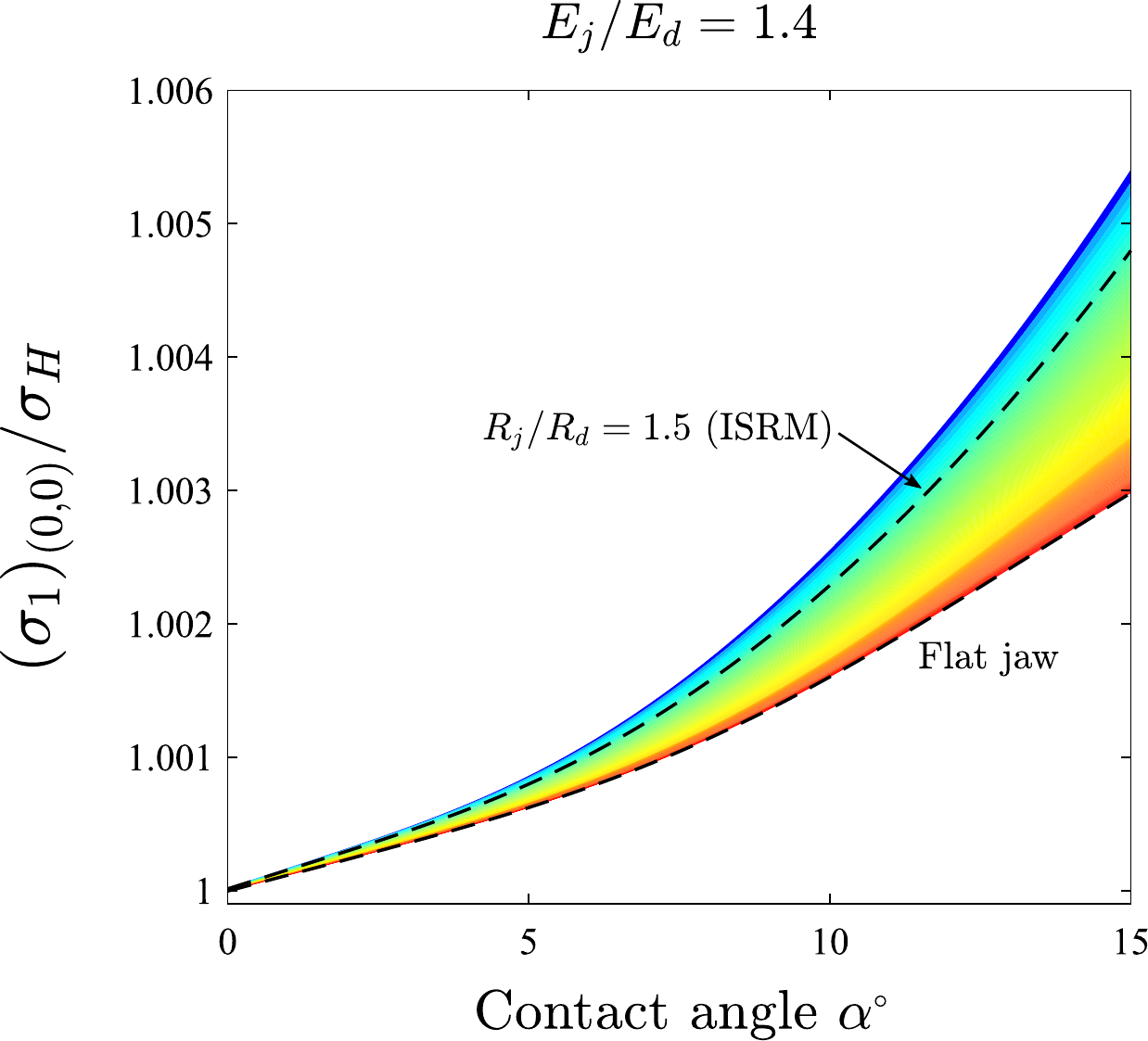}
    \caption{}
    \label{fig:H-E-Constant-b}
    \end{subfigure}\hspace{1 mm}
    \begin{subfigure}[t]{0.088\textwidth}\vspace{-68 mm}
    \includegraphics[width=\textwidth]{legend_R.pdf}
    \end{subfigure}
    \caption{Maps to evaluate the accuracy of Hondros's analytical solution for a uniformly distributed load, Eq. (\ref{eq:Hondros2}). The finite element predictions of the maximum principal stress at the disk centre are normalised by Hondros's stress solution, denoted as $\sigma_H$. The results are obtained for a wide range of $R_j/R_d$ values and the following bounds of the elastic stiffness: (a) $E_j/E_d=42$, and (b) $E_j/E_d=1.4$.}
    \label{fig:HE-Constant}
\end{figure}

Let us assume that the contact angle can be experimentally determined and assess the accuracy of Hondros's analytical solution for $\alpha > 0$, Eq. (\ref{eq:Hondros2}). The finite element prediction of maximum principal stress at the disk centre is shown in Fig. \ref{fig:HE-Constant} normalised by Hondros's analytical solution for a uniformly distributed load. Results are shown for the lower and upper bounds of the elastic modulus considered above, and as a function of the jaw radius. Differences are overall small, as could be expected from Saint-Venant's principle. However, the assumption of a uniform pressure, intrinsic to Hondros's solution, leads to errors above 3\% for softer rocks and curved jaw configurations such as that of the ISRM standard. As in Figs. \ref{fig:load-R-Constant} and \ref{fig:load-E-Constant}, the error becomes negligible for rocks on upper end of the stiffness spectrum and for jaws with large radius. Notwithstanding, as discussed below, the use of a large jaw radius favours the nucleation of cracking far from the disk centre, making the test invalid. 

\subsection{Mapping the conditions that lead to cracking at the disk centre}
\label{sec:MapsCentre}

Low contact angles lead to stress states that are close to the Hondros equations. However, this is not sufficient for the test to be valid as cracking can nucleate outside of the disk centre, as it is often reported when flat or large-radius jaws are used (see, e.g., \cite{Bouali2021,Lin2014,Yuan2017}). While the maps presented in Section \ref{Sec:AnalysisStressCentre} provide a relationship between the critical load and the tensile strength (even if Eq. (\ref{eq:S1withLoad}) is not met), this is only meaningful if the critical load is associated with the initiation of cracks at the disk centre and not elsewhere. To determine the location of crack nucleation, we combine the generalised Griffith failure envelope and finite element analysis (see Section \ref{Sec:GriffithBrazilian}). To achieve this, we start by assuming that cracking initiates at the disk centre, where $\sigma_1=\sigma_t$, and assess that assumption by comparing the compressive-to-tensile strength ratio resulting from the test with the admissible range of $\sigma_c/\sigma_t$ ratios. If the latter is greater than the former, then cracking initiates outside of the disk centre and the test is invalid. Specifically, for each combination of material and test parameters, the process is as follows. Firstly, a finite element analysis is conducted to estimate the principal stresses ($\sigma_1$, $\sigma_3$) at each integration point for a wide range of load increments. Secondly, Eq. (\ref{Eq:G-Com}) is used to compute the minimum admissible $\sigma_c$ (i.e., the maximum $\sigma_c$ among all material points). Finally, from the threshold $\sigma_c$ and the assumption $(\sigma_1)_{(0,0)}=\sigma_t$, a data point is established relating the material and test parameters to the threshold of admissible $\sigma_c/\sigma_t$ values. Each map, such as Fig. \ref{fig:T-Ed}a, is built using approximately 20,000 of these data points and interpolating in-between. The process is automated by means of Python and MATLAB scripts \cite{AES2017}. 

\subsubsection{The influence of the jaw radius}

\begin{figure}[H]
    \centering
    \begin{subfigure}[t]{0.45\textwidth}
    \includegraphics[width=\textwidth]{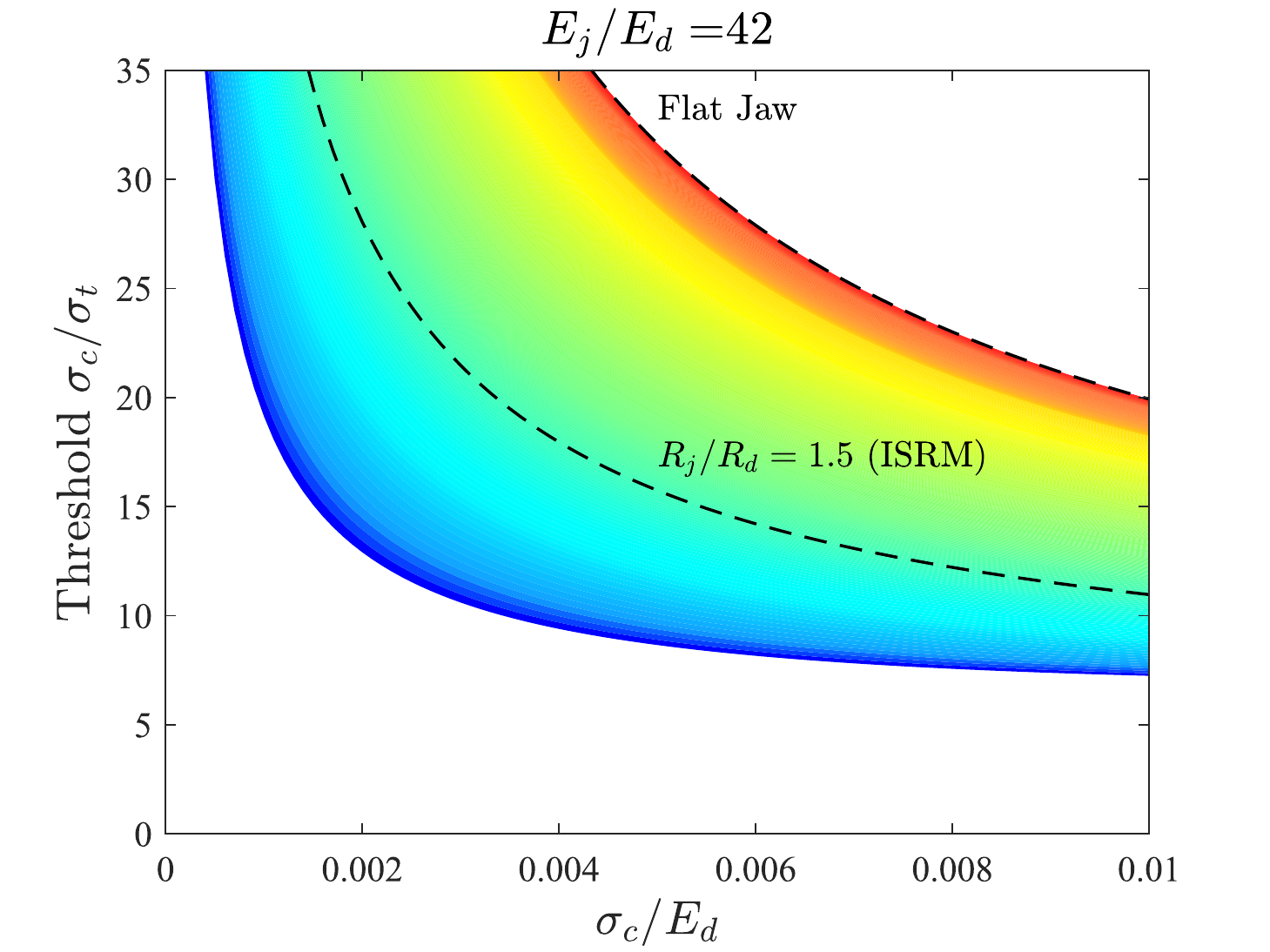}
    \caption{}
    \label{fig:T-Ed-a}
    \end{subfigure}
    \begin{subfigure}[t]{0.45\textwidth}
    \includegraphics[width=\textwidth]{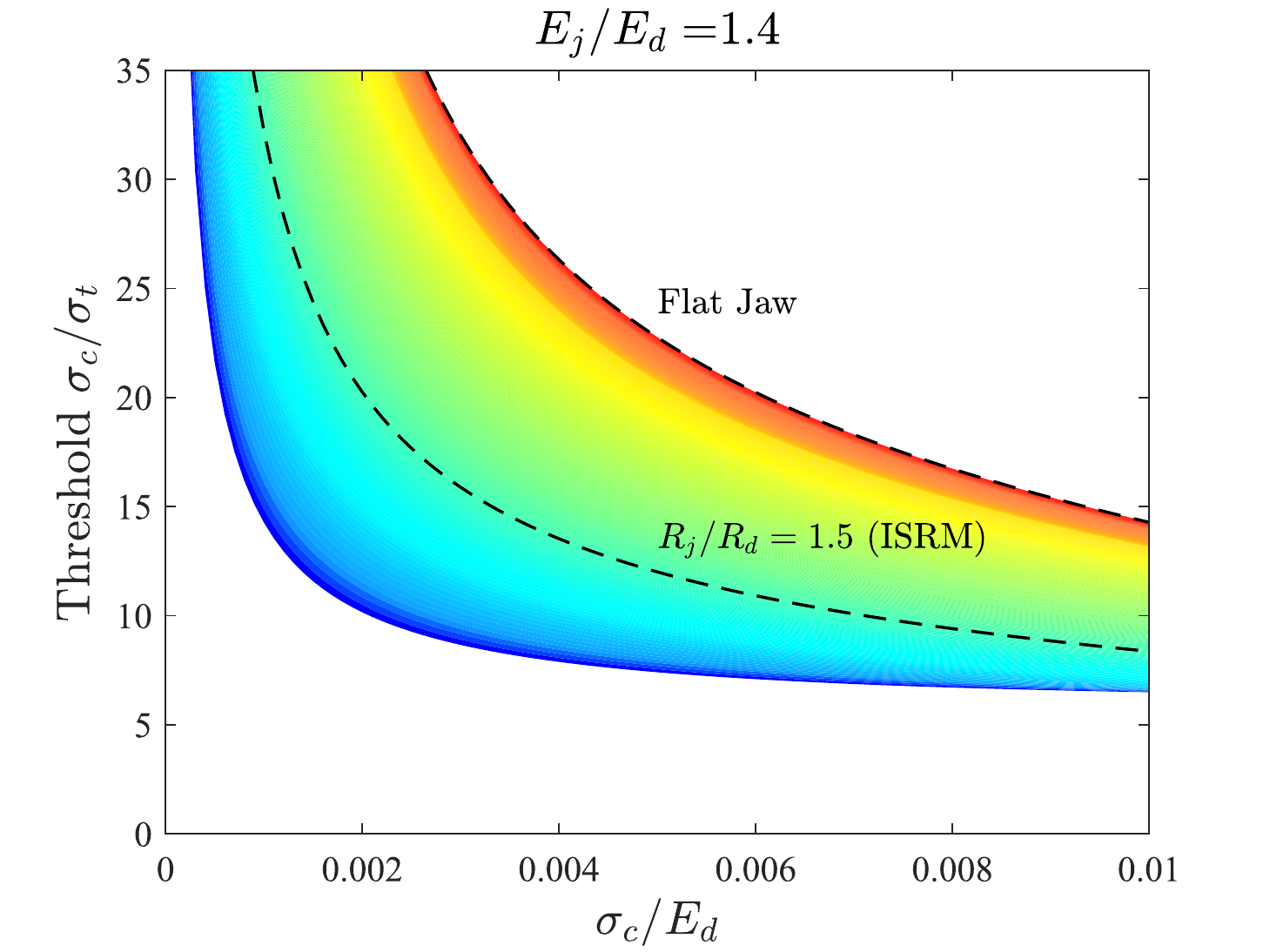}
    \caption{}
    \label{fig:T-Ed-b}
    \end{subfigure}
    \begin{subfigure}{0.079\textwidth}\vspace{-56.5 mm}
    \includegraphics[width=\textwidth]{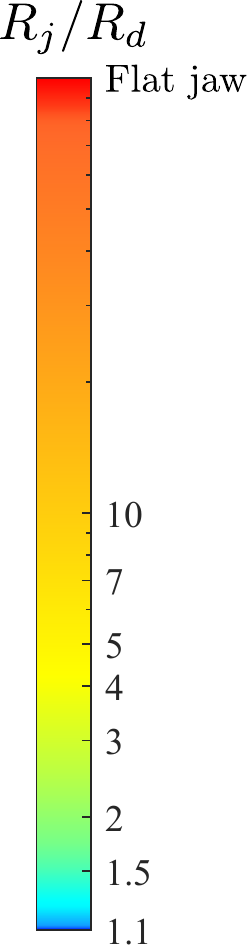}
    \label{}
    \end{subfigure}
    \caption{Maps to assess if cracking nucleates at the centre. Influence of the jaw radius on the minimum acceptable ratio of compressive-to-tensile strength for (a) $E_j/E_d=42$ and (b) $E_j/E_d=1.4$. The disk's Poisson's ratio equals $\nu_d=0.2$. Dashed lines are used to define the conditions relevant to the ASTM \cite{ASTMD3697} and ISRM \cite{Bieniawski1978} standards.}
    \label{fig:T-Ed}
\end{figure}

We start by mapping the influence of the jaw radius on the validity of the Brazilian test. Fig. \ref{fig:T-Ed} shows, following the procedure described above, the relation between the jaw radius (as given by $R_j/R_d$), the non-dimensional set $\sigma_c/E_d$ and the minimum acceptable compressive-to-tensile strength ratio. Maps are provided for two limit cases of disk elastic properties: $E_j/E_d=42$ (i.e., $E_d \approx 5$ GPa) and $E_j/E_d=1.4$ (i.e., $E_d \approx 150$ GPa), with the majority of rock-like materials expected to fall between these two cases. By comparing Figs. \ref{fig:T-Ed}a and \ref{fig:T-Ed}b, it can be seen that while $E_j/E_d$ influences the results, the role appears to be of secondary nature relative to the influence of the jaw radius. 

The results reveal the following trends. First, for a given jaw radius, the range of admissible $\sigma_c/\sigma_t$ ratios increases with increasing $\sigma_c/E_d$, as valid tests (centre cracking) are those above the $\sigma_c/\sigma_t$ threshold. When the compressive strength increases, the likelihood of cracking nucleating outside of the disk centre decreases. For example, consider the specific case $E_j/E_d = 42$ and $R_j/R_d = 1.5$. When $\sigma_c/E_d = 0.002$, the region of validity is $\sigma_c/\sigma_t>28$, whereas when $\sigma_c/E_d = 0.01$, the ratio $\sigma_c/\sigma_t$ needs only to exceed 11. Also, lower $E_d$ values result in larger contact angles and thus less chances of cracking occurring nearby the loading jaws. This is also observed by comparing Figs. \ref{fig:T-Ed}a and \ref{fig:T-Ed}b; the stiffer the sample the more likely that cracking will occur in the compressive regions. Importantly, the results provide $\sigma_c/\sigma_t$ thresholds below which it is not possible to obtain a valid Brazilian test. Thus, it is not possible to obtain a valid result if $\sigma_c/\sigma_t<7$, independently of the jaw radius. For $\sigma_c/E_d$ ratios as high as 0.01, the ISRM ($R_j/R_d=1.5$) configuration provides thresholds of $\sigma_c/\sigma_t$ equal to 11 (Fig. \ref{fig:T-Ed}a) and 8 (Fig. \ref{fig:T-Ed}b). While the ASTM (flat jaws) configuration gives $\sigma_c/\sigma_t$ thresholds of 20 (Fig. \ref{fig:T-Ed}a) and 14 (Fig. \ref{fig:T-Ed}b). Hence, as it can be seen in Fig. \ref{fig:fcE-fcft-total}, conducting Brazilian tests in agreement with the ISRM and (particularly) ASTM guidelines will lead to invalid results for a range of rocky materials, independently of the jaw radius.

\subsubsection{The influence of Young's modulus}

\begin{figure}[H]
    \centering
    \begin{subfigure}{0.49\textwidth}
    \includegraphics[width=\textwidth]{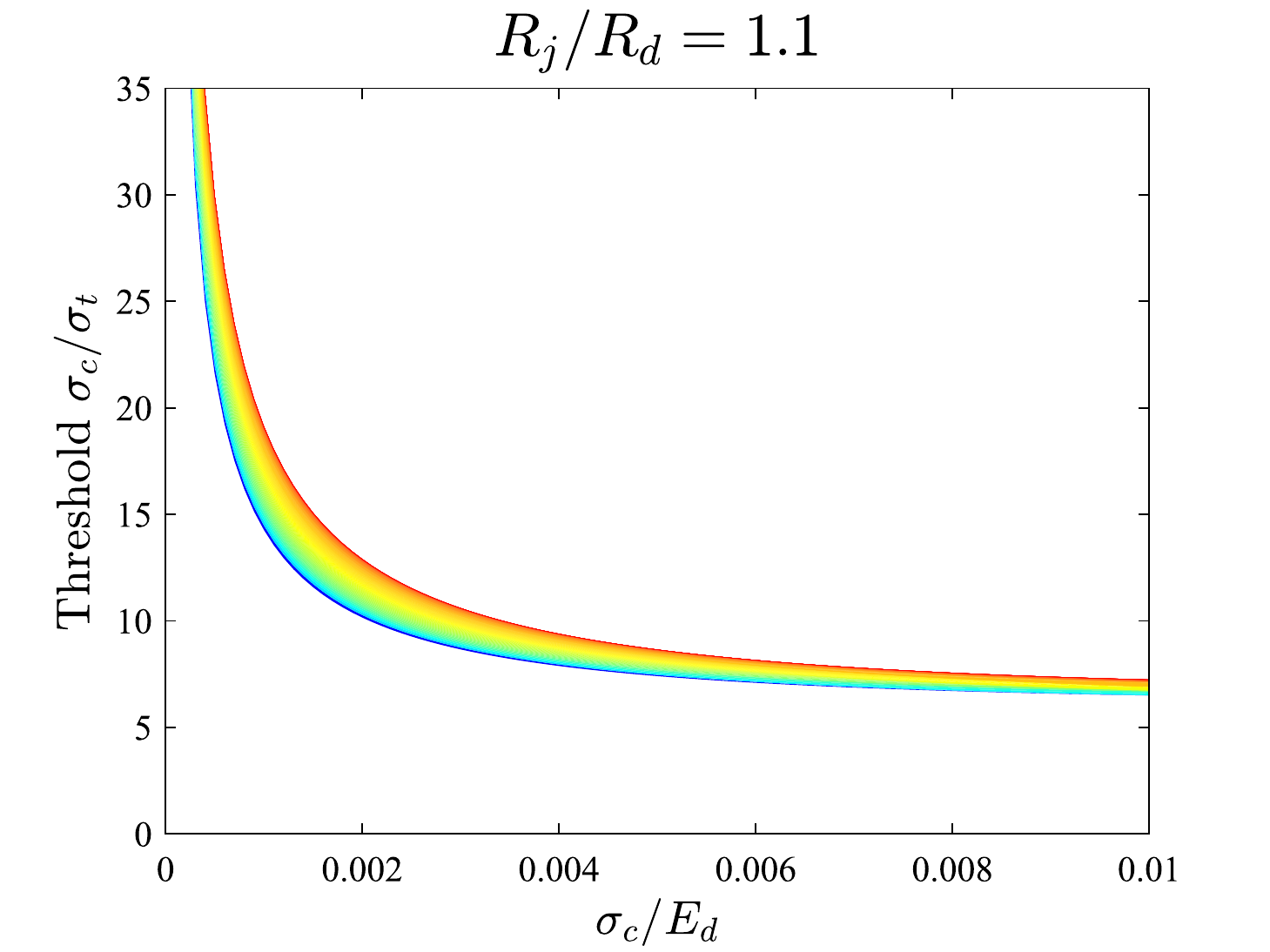}
    \caption{}
    \label{fig:T-Rd-a}
    \end{subfigure}
    \begin{subfigure}{0.49\textwidth}
    \includegraphics[width=\textwidth]{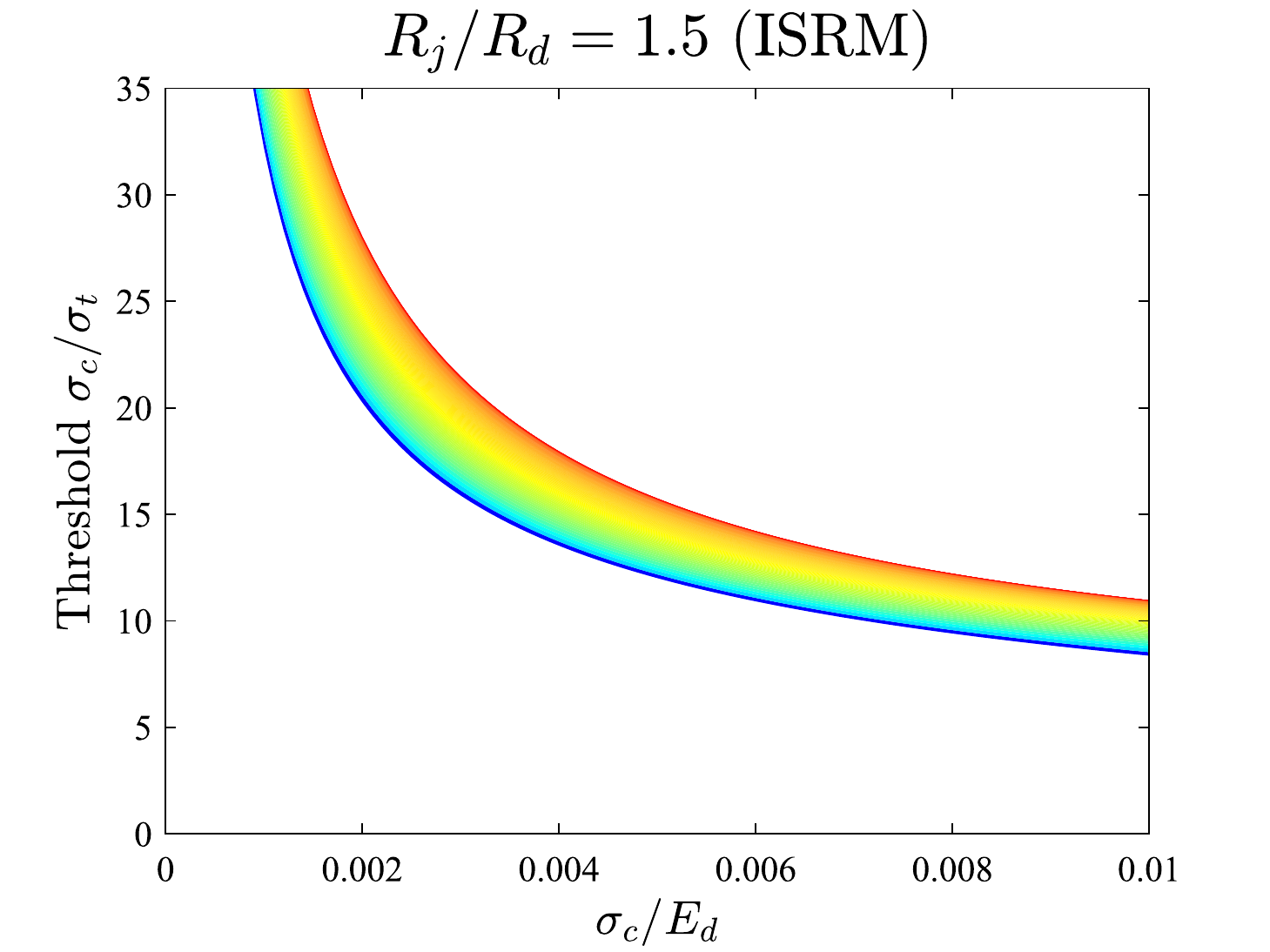}
    \caption{}
    \label{fig:T-Rd-b}
    \end{subfigure}\vspace{5 mm}
    \begin{subfigure}{0.49\textwidth}
    \includegraphics[width=\textwidth]{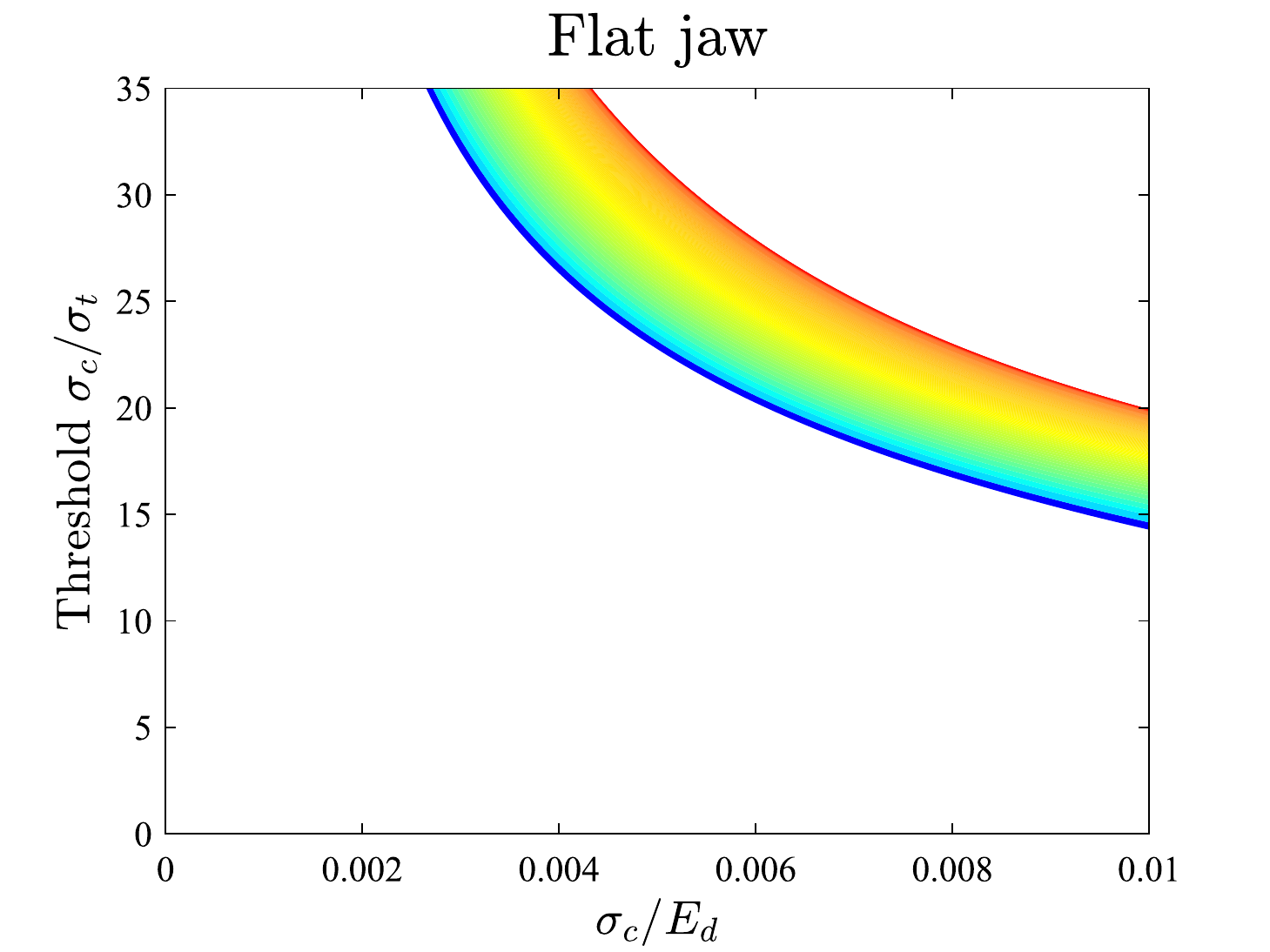}
    \caption{}
    \label{fig:T-Rd-c}
    \end{subfigure}
    \begin{subfigure}{0.0505\textwidth}\vspace{-5 mm}
    \includegraphics[width=\textwidth]{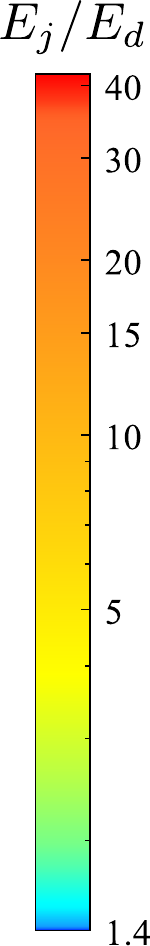}
    \label{}
    \end{subfigure}
    \caption{Maps to assess if cracking nucleates at the centre. Influence of the elastic modulus of the material on the minimum acceptable ratio of compressive-to-tensile strength for (a) $R_j/R_d=1.1$ (a low jaw radius), (b) $R_j/R_d=1.5$ (the ISRM configuration), and (c) flat jaws (the ASTM configuration). The disk's Poisson's ratio equals $\nu_d=0.2$.}
    \label{fig:T-Rd}
\end{figure}

We proceed to report the effect of the Young's modulus of the sample ($E_d$) for selected testing geometries. Specifically, results are shown for a small jaw radius ($R_j/R_d=1.1$) and the ISRM ($R_j/R_d=1.5$) and ASTM (flat jaws) recommended configurations. The maps obtained are presented in Fig. \ref{fig:T-Rd}.\\

Several observations can be drawn. First, the flatter the jaws the higher the sensitivity to the elastic stiffness of the sample. The map is wider and more significant differences can be observed between the admissible limits for a given $\sigma_c/E_d$ value. A smaller range of admissible $\sigma_c/\sigma_t$ ratios (i.e., lower threshold values) is predicted with increasing jaw radius. This is consistent with expectations in terms of contact angles; high contact angles can readily be achieved with curved jaws while flat or large radius jaws can only do so if the disk is soft. Second, the figure emphasises the limitations of current standardised procedures. As shown in Fig. \ref{fig:fcE-fcft-total}, many materials lie within the region delimited by 0.001-0.004 $\sigma_c/E_d$ and 5-15 $\sigma_c/\sigma_t$. However, the maps obtained for the ISRM and ASTM standards fall above this region, implying that the tests will necessarily result in estimates below the admissible $\sigma_c/\sigma_t$ threshold and thus cracking is predicted to occur in the compressive region, rather than in the disk centre. 

\subsubsection{The influence of Poisson's ratio}

\begin{figure}[H]
    \centering
    \includegraphics[width=0.6\textwidth]{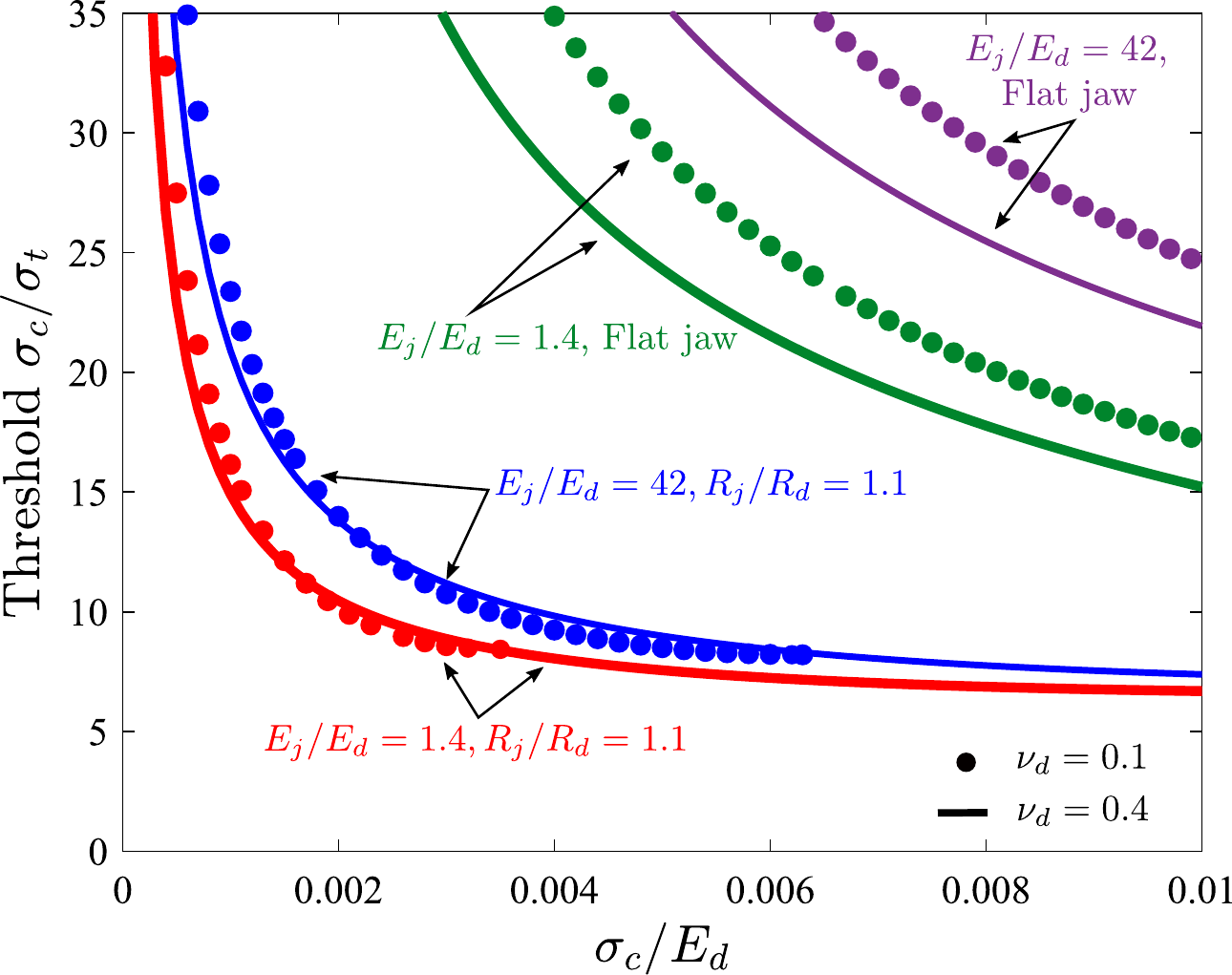}
    \caption{Maps to assess if cracking nucleates at the centre. Influence of the Poisson ratio of the material on the minimum acceptable ratio of compressive-to-tensile strength. Results are obtained for the lower and upper bounds of $\nu_d$ (0.1, 0.4), $E_j/E_d$ (1.4, 42) and $R_j/R_d$ (1.1, 100).}
    \label{fig:PoissonEffect}
\end{figure}

The role of the disk's Poisson's ratio is examined in Fig. \ref{fig:PoissonEffect}. Two limit values are considered, $\nu_d=0.1$ and $\nu_d=0.4$, and results are obtained for limit cases of $E_j/E_d$ and $R_j/R_d$ so as to span all scenarios. Overall, Poisson's ratio seems to play a very secondary role. The effect is negligible for low jaw radii ($R_j/R_d=1.1$) and this appears to be insensitive to the elastic modulus mismatch ($E_j/E_d$). Some differences are observed for jaws with a large radius, with smaller Poisson's ratios further reducing the range of admissible compressive-to-tensile strength ratios. This implies that the appropriate value of Poisson's ratio must be used when assessing the validity of the Brazilian test in a configuration with flat or large-radius jaws, as in the ASTM standard \cite{ASTMD3697}.

\subsubsection{The influence of friction}
\label{Sec:Effect of friction}

\begin{figure}[H]
    \centering
    \includegraphics[width=0.7\textwidth]{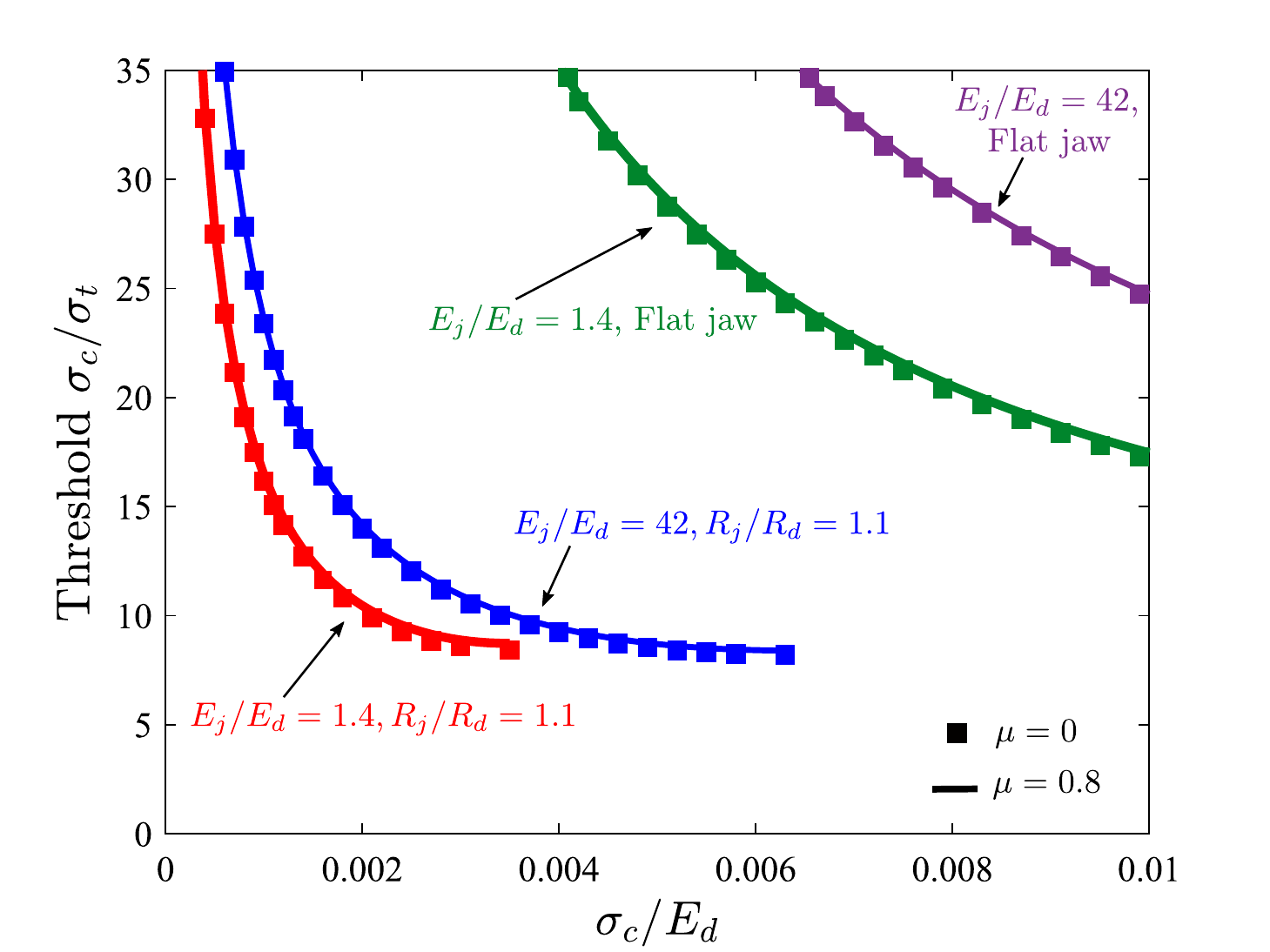}
    \caption{Maps to assess if cracking nucleates at the centre. Influence of friction on the minimum acceptable ratio of compressive-to-tensile strength. Results are obtained without friction and for a friction coefficient of $\mu=0.8$, for the lower and upper bounds of $E_j/E_d$ (1.4, 42) and $R_j/R_d$ (1.1, 100). The disk's Poisson's ratio equals $\nu_d=0.1$.}
    \label{fig:friction effect}
\end{figure}

To investigate the role of friction, simulations are conducted with a friction coefficient of $\mu=0.8$, an upper bound with respect to the values that may be expected for rock/metal interfaces. A penalty method is used to incorporate friction into the model. As in the Poisson's ratio study, we consider limit values of $E_j/E_d$ and $R_j/R_d$, to span all relevant conditions. The results are shown in Fig. \ref{fig:friction effect} for a Poisson's ratio of $\nu_d=0.1$; consistent with the observations above, other values of the disk's Poisson's ratio led to identical conclusions. As it can be observed, no noticeable differences are seen between the simulations with and without friction. This also holds for other values of the friction coefficient (results not shown) and is in agreement with the secondary role of friction reported in the literature \cite{Lavrov2002,Markides2011,Markides2012,Markides2013}. While friction is known to influence the stress state of material points near the jaws \cite{Hooper1971}, these points appear to play a secondary role in our analysis of the validity of the Brazilian test.

\subsection{Representative case studies}
\label{Sec:CaseStudiesMapRock}

\begin{figure}[H]
    \centering
    \begin{subfigure}[t]{0.47\textwidth}
    \includegraphics[width=\textwidth]{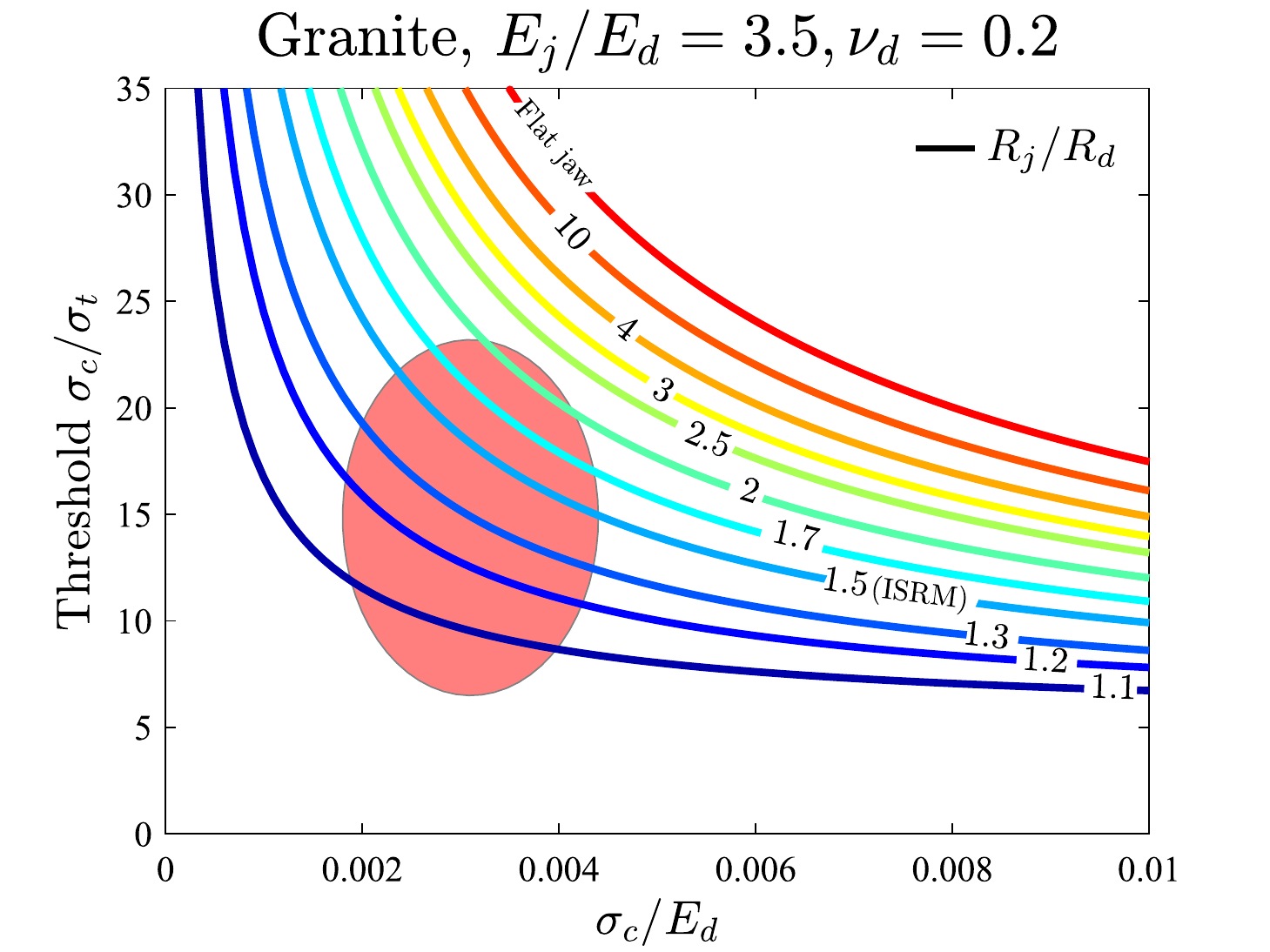}
    \caption{}
    \end{subfigure}
    \begin{subfigure}[t]{0.47\textwidth}
    \includegraphics[width=\textwidth]{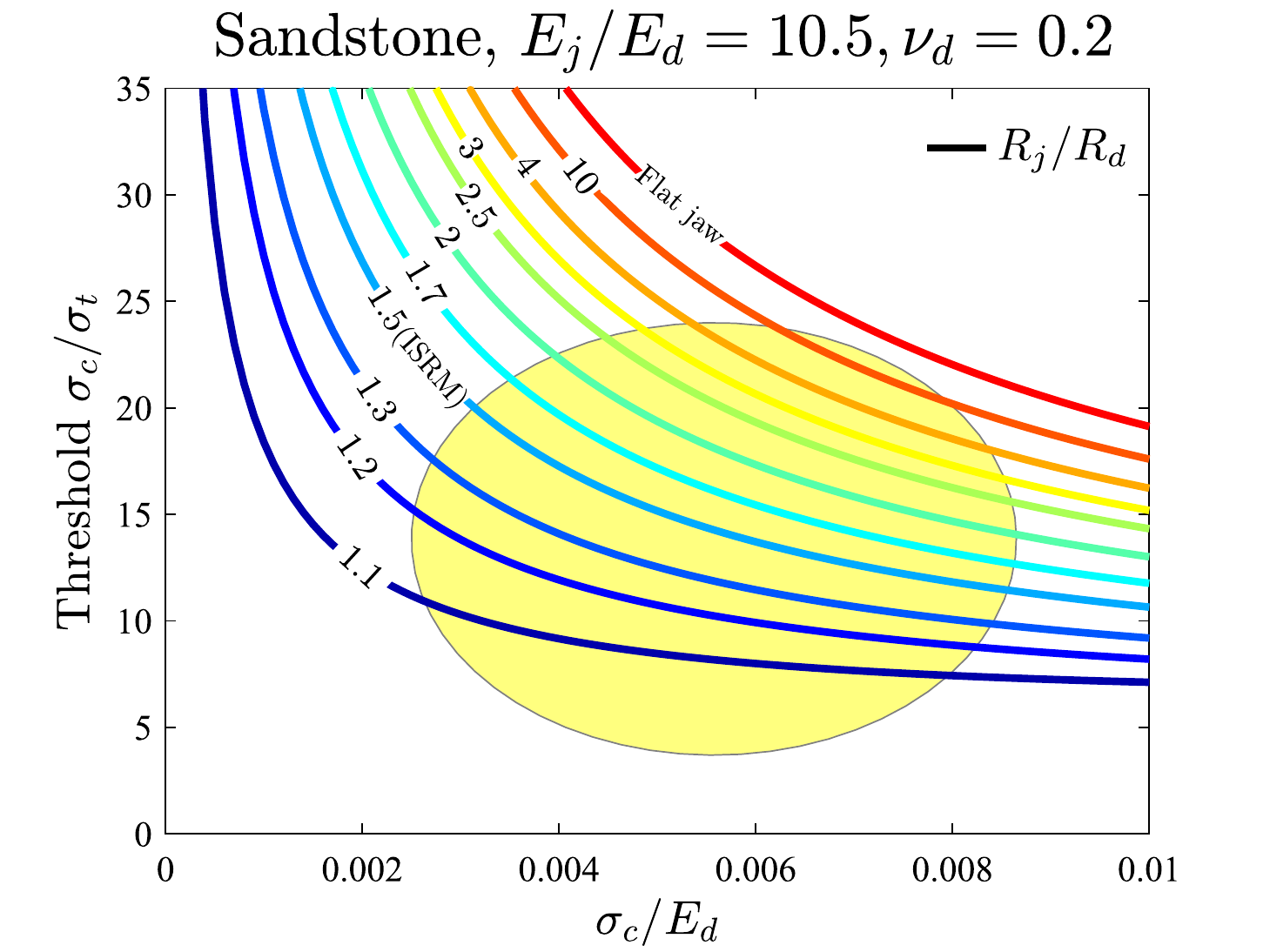}
    \caption{}
    \end{subfigure}
    \begin{subfigure}[t]{0.47\textwidth}
    \includegraphics[width=\textwidth]{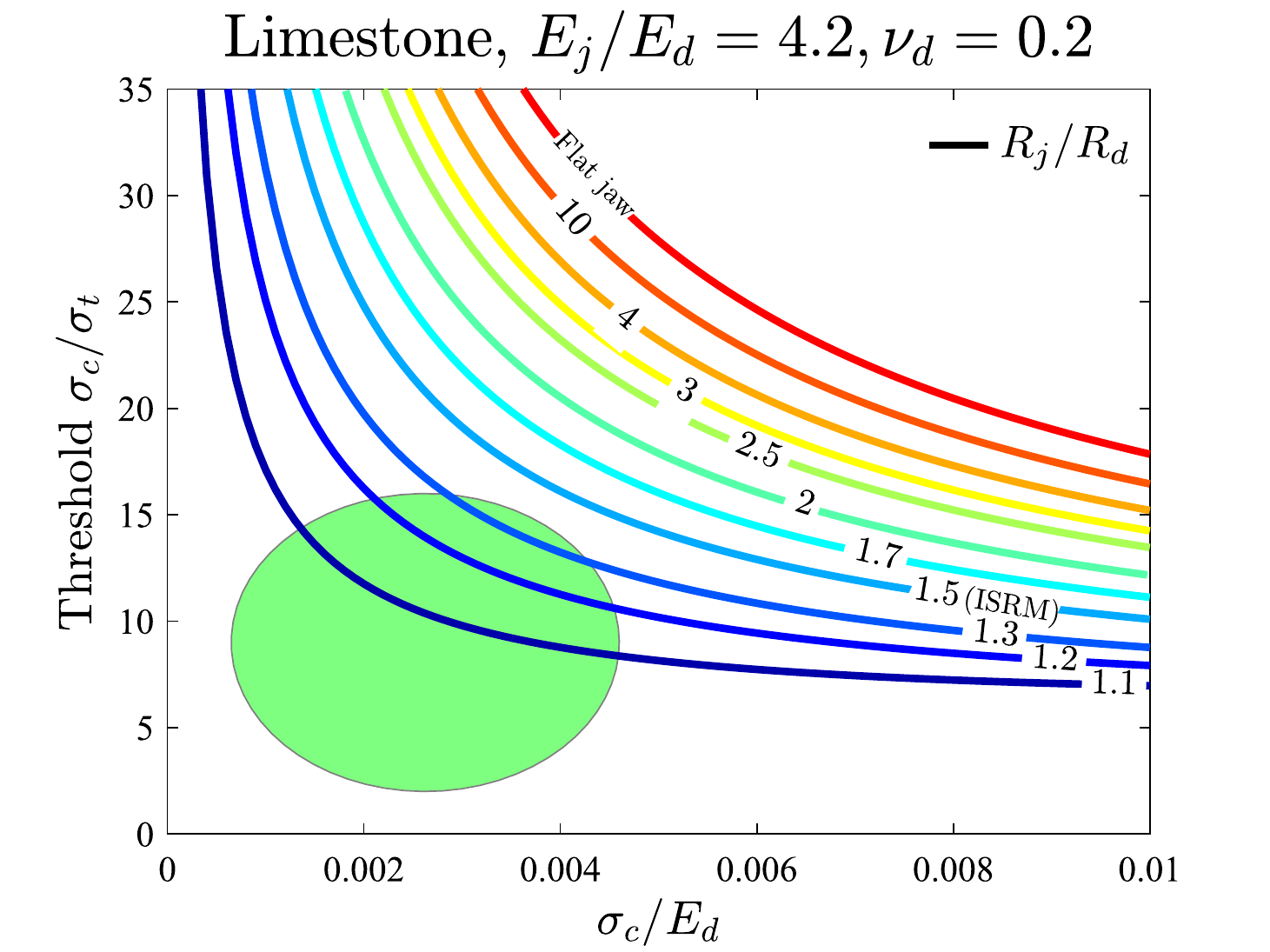}
    \caption{}
    \end{subfigure}
    \begin{subfigure}[t]{0.47\textwidth}
    \includegraphics[width=\textwidth]{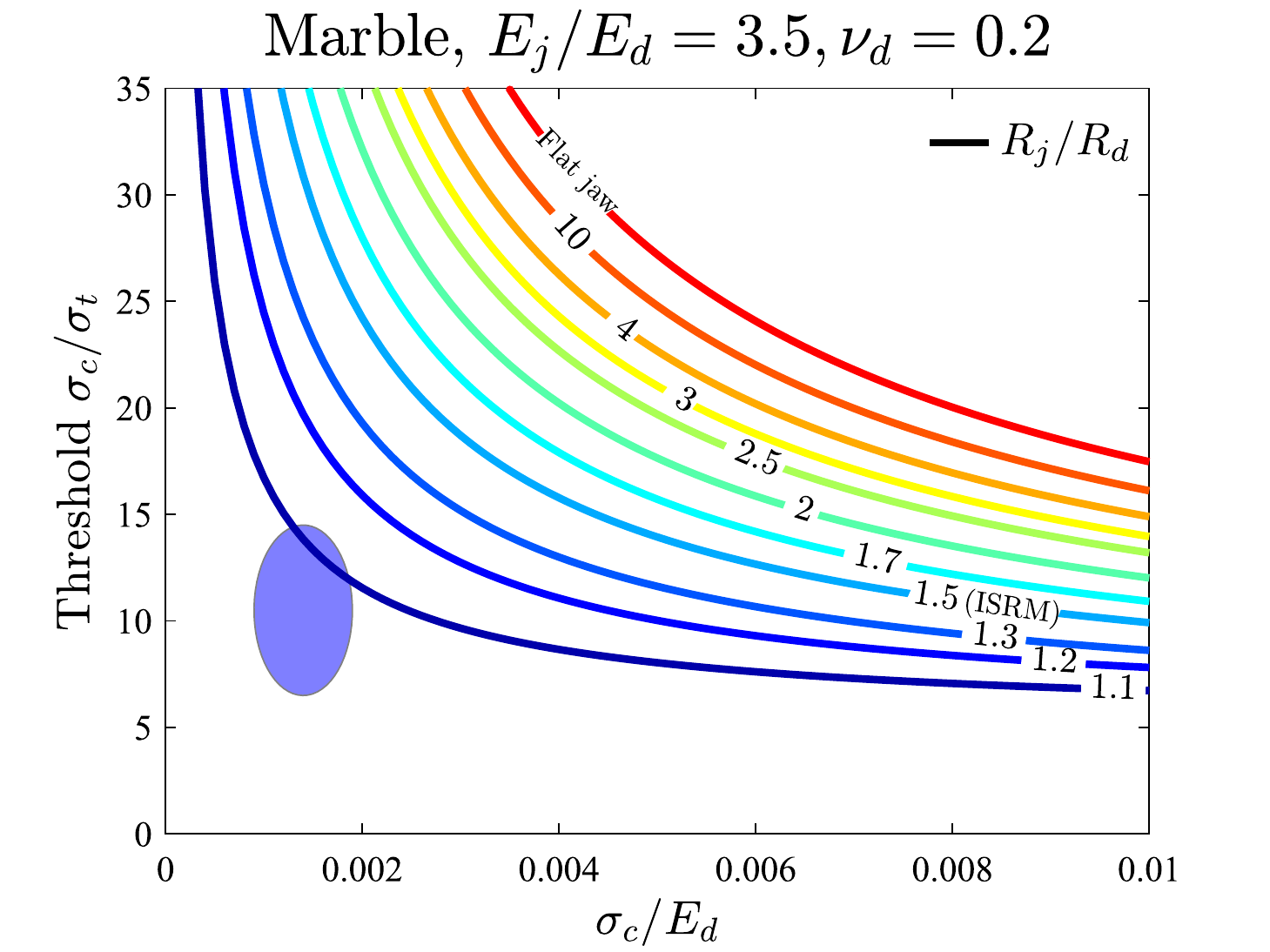}
    \caption{}
    \end{subfigure}
    \caption{Maps to assess if cracking nucleates at the centre: application to: (a) granite, (b) sandstone, (c) limestone, and (d) marble. The figure shows admissible compressive-to-tensile strength ratios as a function of the jaw radius ($R_j/R_d$) for the material properties of: (a) granite ($E_j/E_d=1.5$, $\nu_d=0.2$), (b) sandstone ($E_j/E_d=10.5$, $\nu_d=0.2$), (c) limestone ($E_j/E_d=4.2$, $\nu_d=0.2$), and (d) marble ($E_j/E_d=3.5$, $\nu_d=0.2$). Also, the domain of relevance of each material in a $\sigma_c/\sigma_t$ vs $\sigma_c/E_d$ plot is shown superimposed, as extracted from the GRANTA Material library \cite{GRANTA2021}.}
    \label{fig:MapsRockMaterials}
\end{figure}

Let us now showcase the importance of the maps presented above by particularising them to the study of common rock materials. Fig. \ref{fig:MapsRockMaterials} shows the results obtained for granite, sandstone, limestone and marble. To build the maps, a Poisson's ratio of $\nu_d=0.2$ is adopted in all cases, while the Young's modulus equals $E_d=60$ GPa ($E_j/E_d=3.5$) for granite, $E_d=20$ GPa ($E_j/E_d=10.5$) for sandstone, $E_d=50$ GPa ($E_j/E_d=4.2$) for limestone, and $E_d=60$ GPa ($E_j/E_d=3.5$) for marble. The space that these materials occupy in a compressive-to-tensile strength ratio versus $\sigma_c/E_d$ plot is shown by means of ellipses, based on the material properties available in the GRANTA Material library \cite{GRANTA2021} (see Fig. \ref{fig:fcE-fcft-total}). As before, estimates of the admissible $\sigma_c/\sigma_t$ ratios are provided for jaw radii varying from $R_j/R_d=1.1$ to the flat jaws recommended by the ASTM standard \cite{ASTMD3697}.

Consider first the case of granite, Fig. \ref{fig:MapsRockMaterials}a. Flaw radii from $R_j/R_d=1.1$ to $R_j/R_d=2.2$ can be used to obtain valid estimates for granite materials within the upper estimates of compressive-to-tensile strength ratios. This includes the ISRM configuration ($R_j/R_d=1.5$), which appears to be suited for some classes of granite. The number of suitable testing configurations improves for sandstone, see Fig. \ref{fig:MapsRockMaterials}b. Types of sandstone can be adequately tested with jaw radius up to $R_j/R_d=7$ but the use of flat jaws would lead to an invalid result and no testing configuration is suitable for sandstones with low $\sigma_c/\sigma_t$ ratios. In the case of limestone, see Fig. \ref{fig:MapsRockMaterials}c, only jaw radii from $R_j/R_d=1.1$ to $R_j/R_d=1.3$ can be used and these cover only those limestones with high compressive strength. In this case, it is not possible to get a valid estimate of $\sigma_t$ with the ISRM testing configuration for any type of limestone. Finally, the results obtained for marble (Fig. \ref{fig:MapsRockMaterials}d) show that only a small class of marbles can be adequately characterised with the Brazilian test, and this requires using the smallest jaw radius considered ($R_j/R_d=1.1$). Again, as in the case of limestone, it does not appear to be possible to measure the tensile strength of any class of marble using the Brazilian test configuration suggested by the ISRM. Remarkably, the flat jaws recommended by the ASTM standard are shown to be generally unsuited to provide a valid estimate of the tensile strength, across the wide range of granites, sandstones, limestones and marbles considered.\\

The maps presented can be used by experimentalists to assess the validity of the their testing configuration, as described below. To facilitate this, we provide as Supplementary Material admissible $\sigma_c/\sigma_t$ maps for relevant ranges of material properties and testing parameters. Moreover, as described in \ref{App:MATLABapp}, a MATLAB App is provided that includes a convenient graphical user interface to readily confirm the validity of the test, based on the criteria and analyses conducted here.

\section{A protocol for evaluating the validity of the Brazilian test}
\label{Sec:Protocol}

Identifying experimentally the location of crack nucleation in the Brazilian split test is hindered by the brittle behaviour of rocks; theoretical endeavours are needed to map the conditions of validity of the Brazilian test. The generalised Griffith criterion provides a suitable platform to achieve this as its failure envelope is given by two material properties: the tensile strength $\sigma_t$, which is estimated from the Brazilian test, and the compressive strength $\sigma_c$, which can be measured independently. In the following, we use the maps presented in Section \ref{Sec:Results} to provide a protocol to assess the validity of the Brazilian test as a function of the material and testing parameters. This is illustrated with examples of valid and invalid tests taken from the literature.\\

The protocol is a two-step process. First, one has to determine what is the maximum principal stress at the centre of the disk and second, one has to assess if cracking nucleated at the disk centre or elsewhere. Hondros's equations provide an estimate for the first step, but we have seen in Section \ref{Sec:AnalysisStressCentre} that these can be inaccurate. Thus, it is suggested that the maps provided in Section \ref{Sec:AnalysisStressCentre} and in the Supplementary Material are used instead to accurately determine the stress state at the disk centre. This corresponds with the material tensile strength ($\sigma_1=\sigma_t$) if cracking initiated at the centre. The location of crack initiation is assessed by using the maps presented in Section \ref{sec:MapsCentre}; since $\sigma_c$ and $E_d$ are known (they can be measured independently) we can estimate what is the admissible compressive-to-tensile strength ratio $\sigma_c/\sigma_t$ for a choice of jaw radius $R_j/R_d$. If the magnitude of $\sigma_c/\sigma_t$ resulting from the test is below this admissible threshold, then the test is invalid as cracking has nucleated outside of the centre of the disk. Alternatively, one can use this information before the test, using approximate expected values of $\sigma_t$ (e.g., taken from the literature) to decide what is the most suitable testing geometry ($R_j/R_d$).\\

The protocol is exemplified with two examples of valid and invalid tests, taken from the literature. Specifically, we take as case studies the experiments by Sun and Wu \cite{Sun2021d} on sandstone using the ISRM test configuration and the work by Duevel and Haimson \cite{Duevel1997} on granite, also using the ISRM recommended testing geometry. In both cases the jaws were made of steel, with elastic properties $E_j=210$ GPa and $\nu_j=0.3$. For the sandstone tested in Ref. \cite{Sun2021d}, the reported elastic properties are $E_d=19.15$ GPa and $\nu_d=0.17$ and the material compressive strength is $\sigma_c=99.93$ MPa. For the pink Lac du Bonnet granite study by Duevel and Haimson, the elastic properties are given by $E_d=74.2$ GPa and $\nu_d=0.25$, while the compressive strength was found to be $\sigma_c=219$ MPa \cite{Duevel1997,Cai2010}. The Brazilian tests conducted in Ref. \cite{Sun2021d} and Ref. \cite{Duevel1997} led to tensile strengths of $\sigma_t=7.51$ MPa and $\sigma_t=11.4$ MPa, respectively. Following the protocol presented above, we shall start by assessing the stress state at the disk centre at the critical load.\\ 

\begin{figure}[H]
    \centering
    \begin{subfigure}[t]{0.47\textwidth}
    \centering
    \textbf{$\textbf{Sandstone (Sun \& Wu (2021))}$ \\
    ($E_j/E_d=10.96$, $\nu=0.17$)}\par\medskip
    \includegraphics[width=\textwidth]{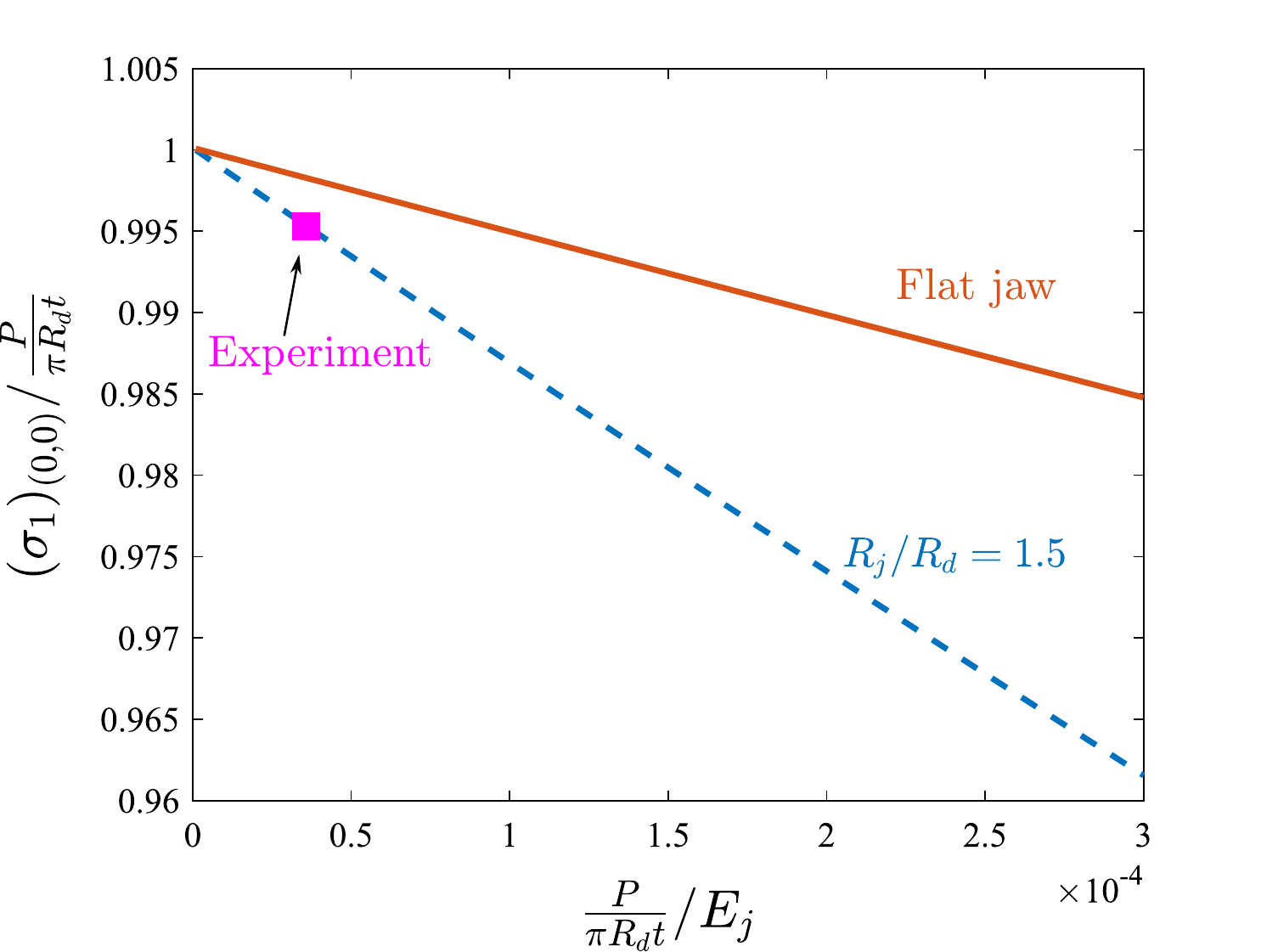}
    \caption{}
    \end{subfigure}
    \begin{subfigure}[t]{0.47\textwidth}
    \centering
    \textbf{$\textbf{Granite (Duevel \& Haimson (1997))}$ \\
    ($E_j/E_d=2.83$, $\nu=0.25$)}\par\medskip
    \includegraphics[width=\textwidth]{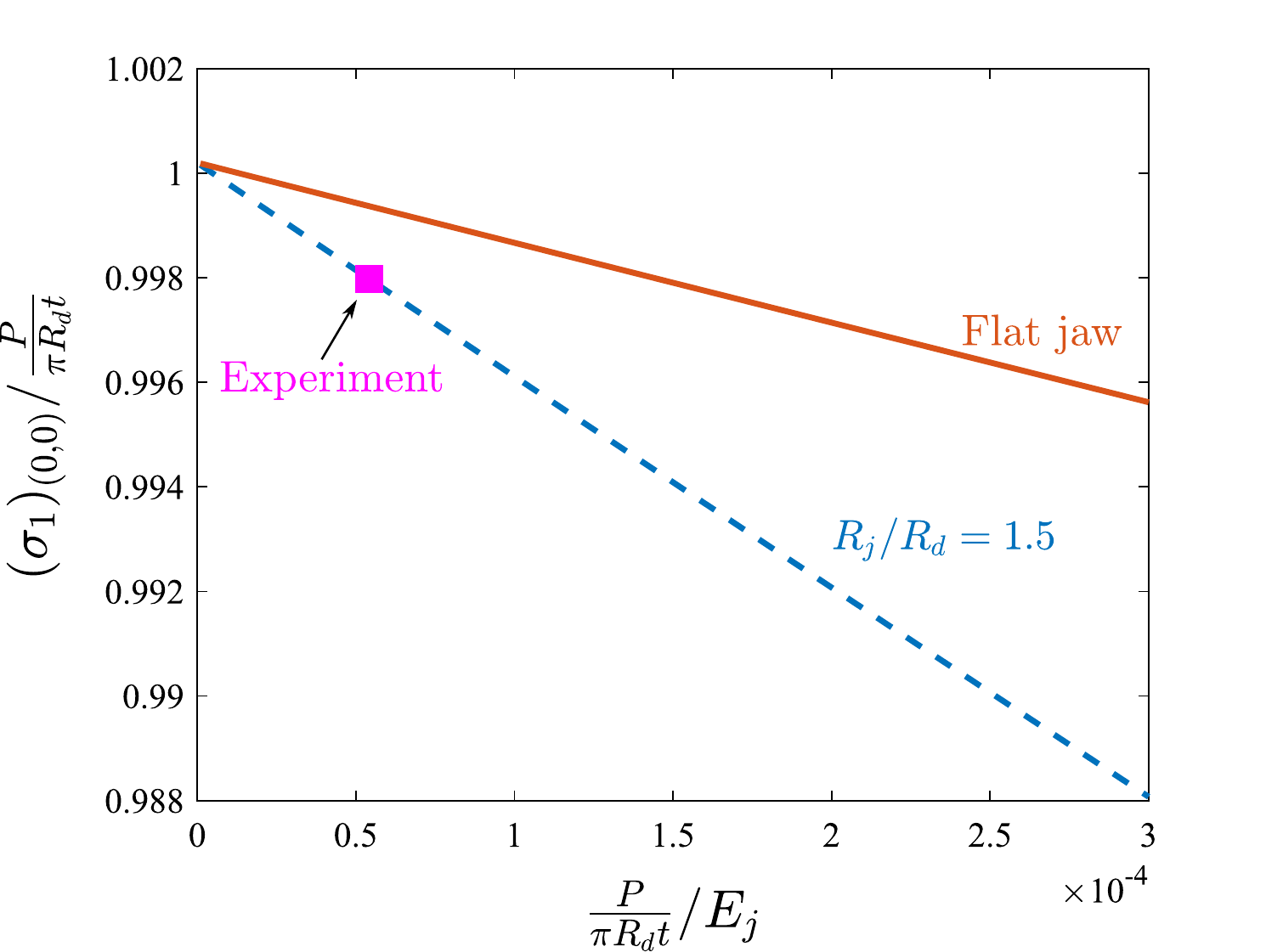}
    \caption{}
    \end{subfigure}
    \caption{A protocol for assessing the validity of the Brazilian test. Step 1 - evaluating the stress state at the disk centre for (a) the sandstone tested by Sun and Wu \cite{Sun2021d}, and (b) the granite tested by Duevel and Haimson \cite{Duevel1997}. The material properties and critical load are $E_j/E_d=10.96$, $\nu_d=0.17$ and $P/(\pi R_d t)=0.0000357E_j$ for (a), and $E_j/E_d=2.83$, $\nu_d=0.25$ and $P/(\pi R_d t)=0.000054E_j$ for (b). The maps provided in Section \ref{Sec:AnalysisStressCentre} and the Supplementary Material are particularised for the two case studies under consideration and the ISRM ($R_j/R_d=1.5$) and ASTM (flat jaws) testing configurations.}
    \label{Fig:CaseStudies_Stress}
\end{figure}

As described above, the first step lies in finding the maximum principal stress $\sigma_1$ at the centre for the critical applied load. Fig. \ref{Fig:CaseStudies_Stress} shows the maps presented in Section \ref{Sec:AnalysisStressCentre} particularised for the two case studies considered here: a sandstone with $E_j/E_d=10.96$ and $\nu_d=0.17$ (Fig. \ref{Fig:CaseStudies_Stress}a) and a granite with $E_j/E_d=2.83$ and $\nu_d=0.25$ (Fig. \ref{Fig:CaseStudies_Stress}b). The results of Fig. \ref{Fig:CaseStudies_Stress} reveal that, while in both case studies the stress state in the disk centre is not described by the point load equation, this approximation provides a good estimate. In the case of the sandstone study by Sun and Wu \cite{Sun2021d} the error relative to Eq. (\ref{eq:S1withLoad}) is below 0.5\% while in the granite experiment by Duevel and Haimson \cite{Duevel1997} the error is roughly 0.2\%. As shown in the figure, a better approximation can be obtained with flat jaws. In any case, Fig. \ref{Fig:CaseStudies_Stress} provides a way of obtaining an accurate estimate of the maximum principal stress at the disk centre, which equals $\sigma_1=7.47$ and $\sigma_1=11.38$ MPa for, respectively, the sandstone and the granite under consideration. These magnitudes correspond to the material tensile strengths, provided that cracking nucleates at the disk centre.\\

\begin{figure}[H]
    \centering
    \begin{subfigure}[t]{0.47\textwidth}
    \centering
    \textbf{$\textbf{Sandstone (Sun \& Wu (2021))}$ \\
    ($E_j/E_d=10.96$, $\nu=0.17$)}\par\medskip
    \includegraphics[width=\textwidth]{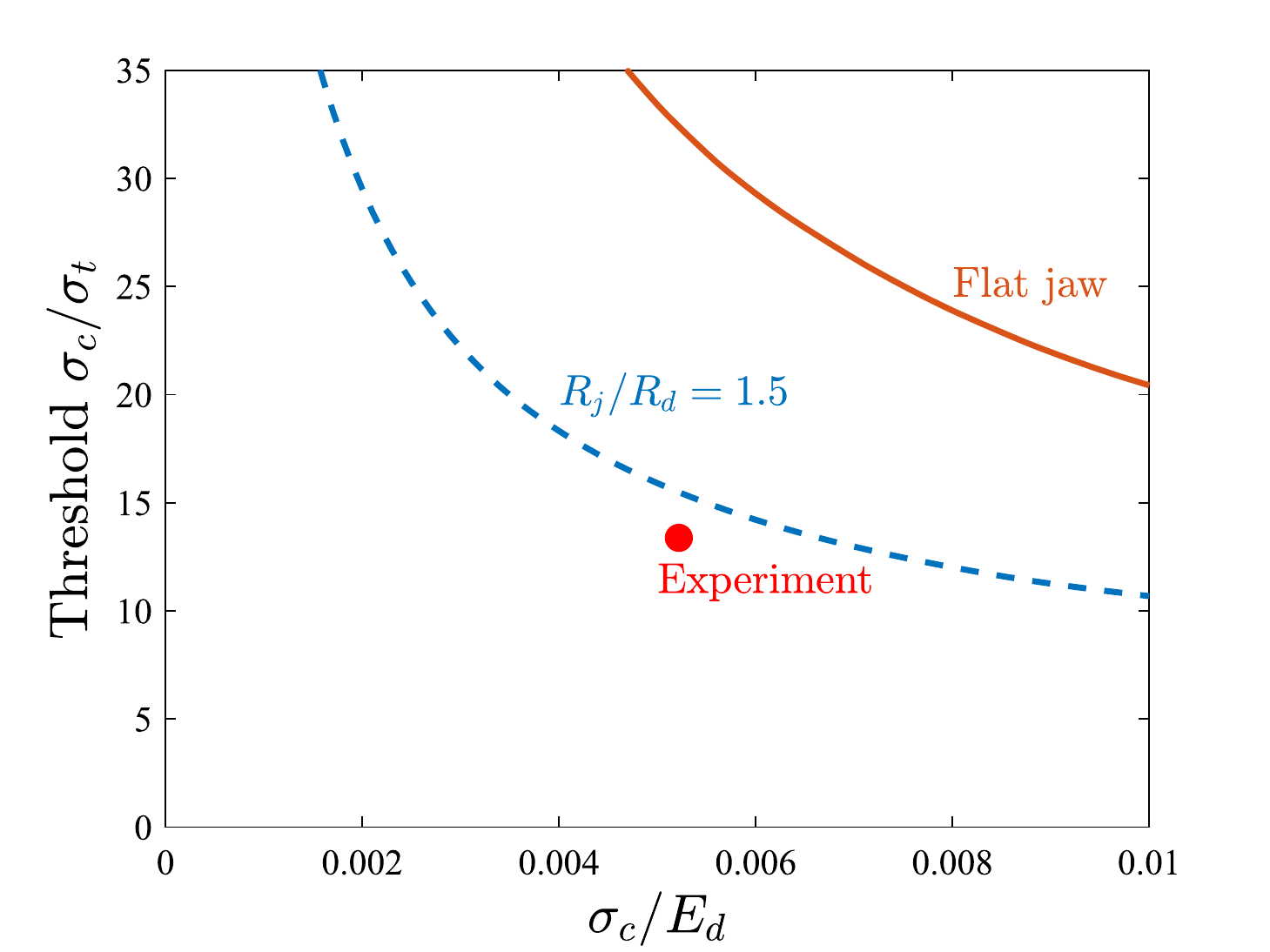}
    \caption{}
    \end{subfigure}
    \begin{subfigure}[t]{0.47\textwidth}
    \centering
    \textbf{$\textbf{Granite (Duevel \& Haimson (1997))}$ \\
    ($E_j/E_d=2.83$, $\nu=0.25$)}\par\medskip
    \includegraphics[width=\textwidth]{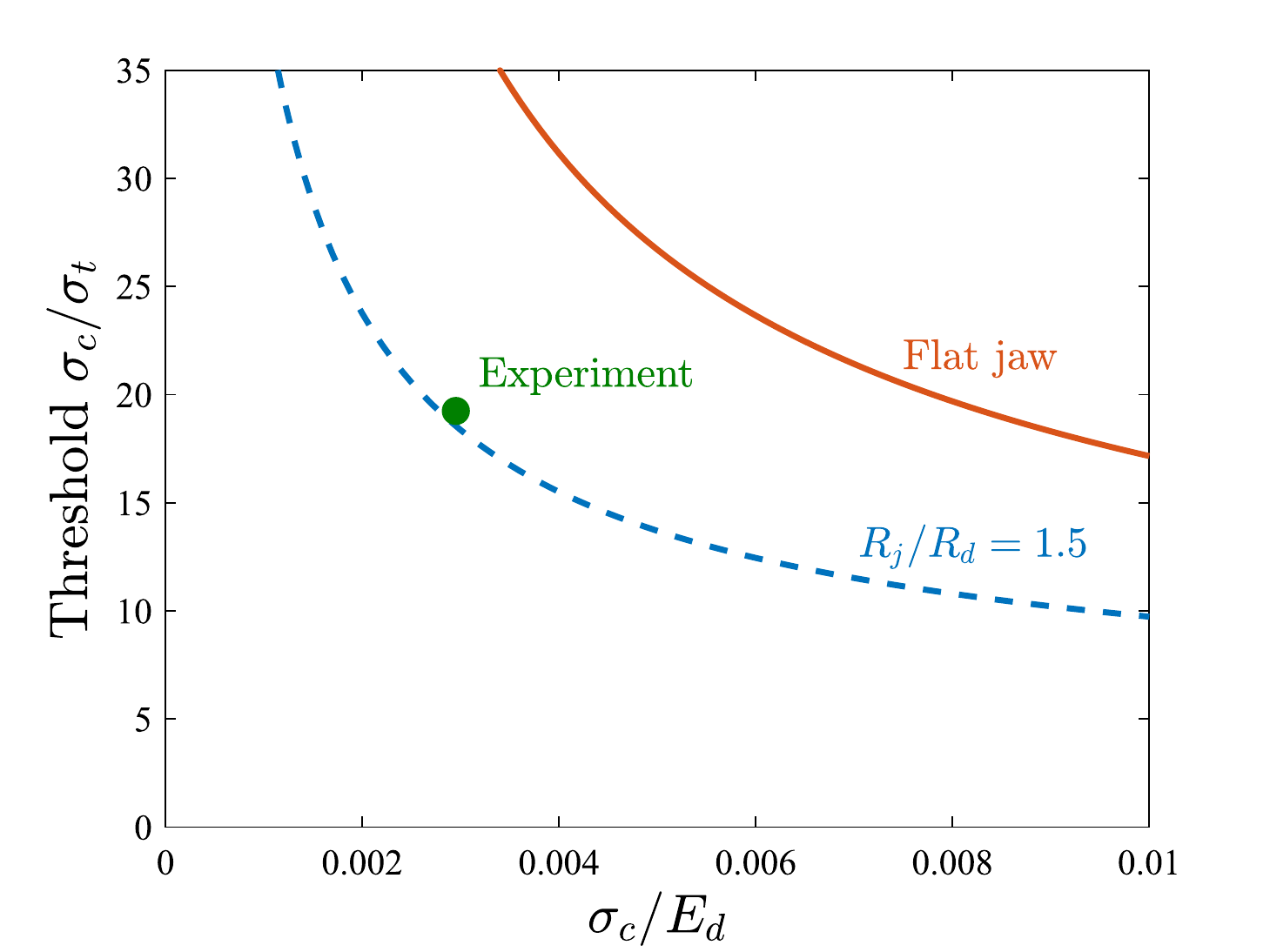}
    \caption{}
    \end{subfigure}
    \caption{A protocol for assessing the validity of the Brazilian test. Step 2 - evaluating the crack nucleation location for (a) the sandstone tested by Sun and Wu \cite{Sun2021d}, and (b) the granite tested by Duevel and Haimson \cite{Duevel1997}. The material properties are $E_j/E_d=10.96$, $\nu_d=0.17$, $\sigma_c/E_d=0.0052$ and $\sigma_c/\sigma_t=13.37$ for (a), and $E_j/E_d=2.83$, $\nu_d=0.25$, $\sigma_c/E_d=0.00295$ and $\sigma_c/\sigma_t=19.24$ for (b). The maps provided in Section \ref{sec:MapsCentre} and the Supplementary Material are particularised for the two case studies under consideration and the ISRM ($R_j/R_d=1.5$) and ASTM (flat jaws) testing configurations. The admissible compression-to-tensile strengths establishes the threshold below which cracking initiates outside of the disk centre and the test becomes invalid.}
    \label{Fig:CaseStudies_Threshold}
\end{figure}

The second and last step involves assessing the crack nucleation location. For the test to be valid, cracking must begin from the disk centre and, following the Griffith's generalised criterion, this will only happen if the compressive-to-tensile strength ratio is above the threshold of admissible values. Thus, given that $\sigma_c$ and $E_d$ are known, we can take the $\sigma_t$ value obtained from the experiment in step 1 and see where the experimental data point lies in the maps presented in Section \ref{sec:MapsCentre}; this is done in Fig. \ref{Fig:CaseStudies_Threshold} for both case studies and the testing geometries recommended by ASTM and ISRM, being the latter the one used in the tests.

The results of Fig. \ref{Fig:CaseStudies_Threshold} show that while the granite study of Duevel and Haimson \cite{Duevel1997} provides a valid estimate of the material tensile strength, this is not the case for the sandstone experiment of Sun and Wu \cite{Sun2021d}. The experimental data point lies below the contour corresponding to the testing geometry employed ($R_j/R_d=1.5$), suggesting that cracking has initiated outside of the centre of the sample. This was also inferred from active and passive ultrasonic techniques in the study by Sun and Wu \cite{Sun2021d}, who concluded that cracking had initiated close to the jaws. Their comprehensive analysis, including numerical and experimental analysis of multiple testing configurations, showcased the limitations of the Brazilian test. The protocol and maps provided here (see also the Supplementary Material and \ref{App:MATLABapp}) enable establishing the conditions where the Brazilian test is valid, upon assuming that crack propagation is well approximated by the generalised Griffith criterion.

\section{Conclusions}
\label{Sec:Conclusions}

We have combined the generalised Griffith criterion and finite element analysis to theoretically assess the validity of the Brazilian split test. Maps have been provided to evaluate, as a function of material properties and test geometry, the fulfilment of the two assumptions inherent to the indirect estimate of the material tensile strength provided by the Brazilian test; that (i) the load is related to the maximum principal stress at the disk centre through Hondros's equations, and that (ii) cracking starts at the centre of the sample. The use of the generalised Griffith criterion enables assessing (ii) using a failure envelope that is solely a function of two material properties that can be independently measured: the tensile ($\sigma_t$) and compressive ($\sigma_c$) strengths. Our main findings are the following:
\begin{itemize}
    \item For relevant contact angles, there is a noticeable deviation from the stress solution for a point load. However, the error remains small (below 5\%) for a wide range of rock-like materials if flat or large-radii jaws are used. 
    \item The use of the Hondros's stress solution for a uniformly distributed load ensures that the error does not exceed 4\% for relevant ranges of stiffness mismatch and jaw radius. However, unlike the maps provided, requires an experimental characterisation of the contact angle at failure. 
    \item The use of jaws with large radii favours the initiation of cracking in the compressive region, far from the disk centre, making the test invalid.
    \item The location of crack initiation is particularly sensitive to the testing geometry and, to a lesser degree, to the stiffness of the sample. Poisson's ratio plays a negligible role in jaws with a small radius but has an effect in the case of flat jaws. No influence of friction is observed.
    \item The analysis of the main classes of rocks reveals that the Brazilian test is not a suitable experiment for a wide range of materials. Only a small set of marbles and limestones (those with high $\sigma_c/\sigma_t$) can be adequately characterised and this requires the use of jaws with small radii. On the other hand, large-radius jaws can be used to test a range of granites and sandstones. The ISRM configuration ($R_j/R_d=1.5$) appears to be solely suitable for these two latter classes of rocks, while the ASTM test geometry (flat jaws) was found to be unsuited to provide a valid estimate of tensile strength for any of the rock-like materials considered.
\end{itemize}

These findings suggest that the regimes of validity of the Brazilian test are much smaller than previously thought. To overcome these shortcomings and determine the range of conditions that lead to a valid Brazilian test, we have provided:
\begin{itemize}
    \item Maps that relate the critical load with the stress state at the disk centre. These allow for accurately estimating the tensile strength without the need of using the approximation provided by the Hondros's equations.
    \item Maps that quantify the admissible compression-to-tensile strength ratios above which cracking initiates at the centre of the disk. These allow determining if the test is valid \textit{a posteriori} or making \textit{a priori} decisions of adequate test geometries based on expected $\sigma_t$ values.
    \item A two-step protocol that will allow experimentalists to determine the validity of the test and accurately estimate the material tensile strength. The protocol is demonstrated with examples of valid and invalid tests from the literature. To facilitate uptake, this is encapsulated into a MATLAB App with an easy user interface.
\end{itemize}

\section{Acknowledgments}
\label{Sec:Acknowledge of funding}

The authors acknowledge financial support from the Ministry of Science, Innovation and Universities of Spain through grant PGC2018-099695-B-I00. E. Mart\'{\i}nez-Pa\~neda additionally acknowledges financial support from the Royal Commission for the 1851 Exhibition through their Research Fellowship programme (RF496/2018).



\appendix

\section{BrazVal: A MATLAB App to assess the validity of the Brazilian test}
\label{App:MATLABapp}

A Matlab App is provided to facilitate the assessment of the validity of the Brazilian test, as per the Griffith generalised criterion and the analysis described in this manuscript. As shown in Fig. \ref{fig:Application}, the MATLAB App contains a simple graphical user interface where the user provides as input variables the parameters related to the disk sample (radius $R_d$, Young's modulus $E_d$, Poisson's ratio $\nu_d$, compressive strength $\sigma_c$ and thickness $t$) and to the jaws (radius $R_j$, Young's modulus $E_j$, Poisson's ratio $\nu_j$), as well as the critical load measured $P_c$. Upon clicking the button \texttt{Run}, the App provides the material tensile strength $\sigma_t$. If the test is deemed invalid, the message \texttt{INVALID} will be shown instead. In addition, the App provides the user with the tensile stress estimate based on Eq. (\ref{eq:S1withLoad}), the actual tensile stress at the disk centre (which will coincide with $\sigma_t$ if the test is valid) and the maximum allowable tensile strength, as determined from the threshold $\sigma_c/\sigma_t$ ratio that ensures that cracking nucleates earlier at the disk centre than elsewhere. 

\setcounter{figure}{0} 

\begin{figure}[H]
    \centering
    \includegraphics[width=.7\textwidth]{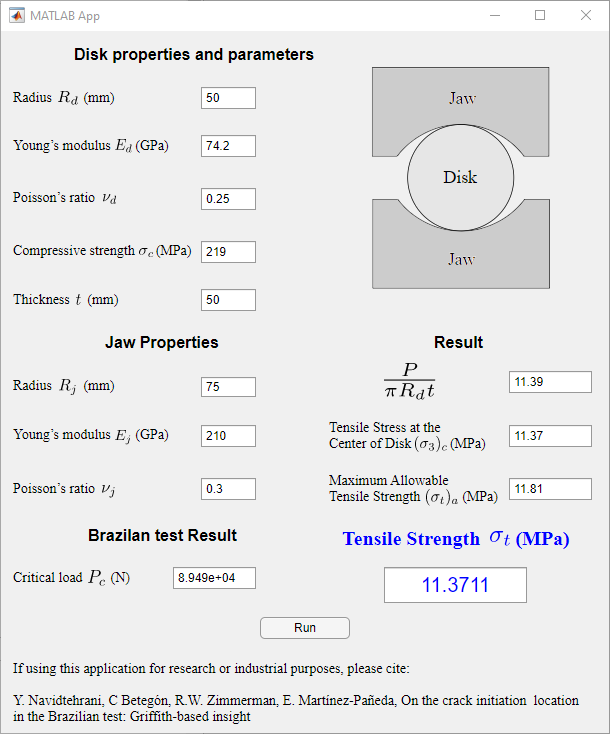}
    \caption{Graphical User Interface (GUI) of BrazVal, a MATLAB App to assess the validity of the Brazilian test, as a function of material and testing parameters. The App can be downloaded from www.empaneda.com/codes.}
    \label{fig:Application}
\end{figure}

The information provided by the App is based on a data grid generated by performing finite element calculations such as those described in Section \ref{Sec:Results}. For scenarios for which data points do not exist, an estimate is attained by using linear interpolation (MATLAB's function \texttt{griddedInterpolant}). The App can be downloaded from www.empaneda.com/codes

\bibliographystyle{elsarticle-num} 
\bibliography{library}

\begin{thebibliography}{10}
\expandafter\ifx\csname url\endcsname\relax
  \def\url#1{\texttt{#1}}\fi
\expandafter\ifx\csname urlprefix\endcsname\relax\def\urlprefix{URL }\fi
\expandafter\ifx\csname href\endcsname\relax
  \def\href#1#2{#2} \def\path#1{#1}\fi

\bibitem{Li2013}
D.~Li, L.~N.~Y. Wong, {The brazilian disc test for rock mechanics applications:
  Review and new insights}, Rock Mechanics and Rock Engineering 46~(2) (2013)
  269--287.

\bibitem{Carneiro1943}
F.~Carneiro, {A new method to determine the tensile strength of concrete}, in:
  Proceedings of the 5th meeting of the Brazilian Association for Technical
  Rules, Vol. Section 3d, 1943.

\bibitem{Akazawa1943}
T.~Akazawa, {New test method for evaluating internal stress due to compression
  of concrete}, Journal of Japan Society of Civil Engineers 29 (1943) 777--787.

\bibitem{Bieniawski1978}
Z.~T. Bieniawski, I.~Hawkes, {Suggested Methods for Determining Tensile
  Strength of Rock Materials}, International Journal of Rock Mechanics and
  Mining Sciences 15~(3) (1978) 99--103.

\bibitem{Hondros1959}
G.~Hondros, {The evaluation of Poisson's ratio and the modulus of materials of
  a low tensile resistance by the Brazilian (indirect tensile) test with
  particular reference to concrete}, Australian Journal of Applied Science
  10~(3) (1959) 243--268.

\bibitem{Timoshenko1951}
S.~Timoshenko, J.~Goodier, {Theory of elasticity}, McGraw-Hill, 1951.

\bibitem{Fairhurst1964}
C.~Fairhurst, {On the validity of the `Brazilian' test for brittle materials},
  International Journal of Rock Mechanics and Mining Sciences and 1~(4) (1964)
  535--546.

\bibitem{Hudson1972}
J.~A. Hudson, E.~T. Brown, F.~Rummel, {The controlled failure of rock discs and
  rings loaded in diametral compression}, International Journal of Rock
  Mechanics and Mining Sciences and 9~(2) (1972) 241--248.

\bibitem{Alvarez-Fernandez2020}
M.~I. Alvarez-Fernandez, C.~C. Garcia-Fernandez, C.~Gonzalez-Nicieza, D.~J.
  Guerrero-Miguel, {Effect of the Contact Angle in the Failure Pattern in Slate
  Under Diametral Compression}, Rock Mechanics and Rock Engineering 53~(5)
  (2020) 2123--2139.

\bibitem{Markides2016}
C.~F. Markides, S.~K. Kourkoulis, {The influence of jaw's curvature on the
  results of the Brazilian disc test}, Journal of Rock Mechanics and
  Geotechnical Engineering 8~(2) (2016) 127--146.

\bibitem{Gutierrez-Moizant2020}
R.~Guti{\'{e}}rrez-Moizant, M.~Ram{\'{i}}rez-Berasategui,
  S.~S{\'{a}}nchez-Sanz, S.~Santos-Cuadros, {Experimental verification of the
  boundary conditions in the success of the Brazilian test with loading arcs.
  An uncertainty approach using concrete disks}, International Journal of Rock
  Mechanics and Mining Sciences 132 (2020) 104380.

\bibitem{Bouali2021}
M.~F. Bouali, M.~Bouassida, {Numerical simulation of the effect of loading
  angle on initial cracks position point: Application to the Brazilian test},
  Applied Sciences 11~(8) (2021).

\bibitem{Garcia-Fernandez2018}
C.~C. Garcia-Fernandez, C.~Gonzalez-Nicieza, M.~I. Alvarez-Fernandez, R.~A.
  Gutierrez-Moizant, {Analytical and experimental study of failure onset during
  a Brazilian test}, International Journal of Rock Mechanics and Mining
  Sciences 103 (2018) 254--265.

\bibitem{Zhao2021}
Z.~Zhao, W.~Sun, S.~Chen, D.~Yin, H.~Liu, B.~Chen, {Determination of critical
  criterion of tensile-shear failure in Brazilian disc based on theoretical
  analysis and meso-macro numerical simulation}, Computers and Geotechnics 134
  (2021) 104096.

\bibitem{Aliabadian2019}
Z.~Aliabadian, G.~F. Zhao, A.~R. Russell, {Failure, crack initiation and the
  tensile strength of transversely isotropic rock using the Brazilian test},
  International Journal of Rock Mechanics and Mining Sciences 122 (2019)
  104073.

\bibitem{Erarslan2012}
N.~Erarslan, Z.~Z. Liang, D.~J. Williams, {Experimental and numerical studies
  on determination of indirect tensile strength of rocks}, Rock Mechanics and
  Rock Engineering 45~(5) (2012) 739--751.

\bibitem{Yu2021}
H.~Yu, D.~H. Andersen, J.~He, Z.~Zhang, {Is it possible to measure the tensile
  strength and fracture toughness simultaneously using flattened Brazilian
  disk?}, Engineering Fracture Mechanics 247 (2021) 107633.

\bibitem{Komurlu2015}
E.~Komurlu, A.~Kesimal, {Evaluation of Indirect Tensile Strength of Rocks Using
  Different Types of Jaws}, Rock Mechanics and Rock Engineering 48~(4) (2015)
  1723--1730.

\bibitem{ASTMD3697}
{ASTM D3697 Standard test method for splitting tensile strength of intact rock
  core specimens}, ASTM International, West Conshohocken, PA.

\bibitem{Griffith1924}
A.~A. Griffith, {The Theory of Rupture}, in: Proc. First International Congress
  for Applied Mechanics, 1924, pp. 55--63.

\bibitem{Jaeger2009}
J.~Jaeger, N.~Cook, R.~Zimmerman, {Fundamentals of rock mechanics}, Blackwell
  Publishing, Oxford, UK, 2009.

\bibitem{Hoek2014}
E.~Hoek, C.~D. Martin, {Fracture initiation and propagation in intact rock - A
  review}, Journal of Rock Mechanics and Geotechnical Engineering 6~(4) (2014)
  287--300.

\bibitem{GRANTA2021}
{Ansys Granta EduPack}, ANSYS Inc., Cambridge, UK, 2021.

\bibitem{Lin2014}
H.~Lin, W.~Xiong, W.~Zhong, C.~Xia, {Location of the Crack Initiation Points in
  the Brazilian Disc Test}, Geotechnical and Geological Engineering 32~(5)
  (2014) 1339--1345.

\bibitem{Yuan2017}
R.~Yuan, B.~Shen, {Numerical modelling of the contact condition of a Brazilian
  disk test and its influence on the tensile strength of rock}, International
  Journal of Rock Mechanics and Mining Sciences 93~(December 2015) (2017)
  54--65.

\bibitem{AES2017}
G.~Papazafeiropoulos, M.~Mu{\~{n}}iz-Calvente, E.~Mart{\'{i}}nez-Pa{\~{n}}eda,
  {Abaqus2Matlab: A suitable tool for finite element post-processing}, Advances
  in Engineering Software 105 (2017) 9--16.

\bibitem{Lavrov2002}
A.~Lavrov, A.~Vervoort, {Theoretical treatment of tangential loading effects on
  the Brazilian test stress distribution}, International Journal of Rock
  Mechanics and Mining Sciences 39~(2) (2002) 275--283.

\bibitem{Markides2011}
C.~F. Markides, D.~N. Pazis, S.~K. Kourkoulis, {Influence of friction on the
  stress field of the brazilian tensile test}, Rock Mechanics and Rock
  Engineering 44~(1) (2011) 113--119.

\bibitem{Markides2012}
C.~F. Markides, D.~N. Pazis, S.~K. Kourkoulis, {The Brazilian disc under
  non-uniform distribution of radial pressure and friction}, International
  Journal of Rock Mechanics and Mining Sciences 50 (2012) 47--55.

\bibitem{Markides2013}
C.~F. Markides, S.~K. Kourkoulis, {Naturally accepted boundary conditions for
  the brazilian disc test and the corresponding stress field}, Rock Mechanics
  and Rock Engineering 46~(5) (2013) 959--980.

\bibitem{Hooper1971}
J.~A. Hooper, {The failure of glass cylinders in diametral compression},
  Journal of the Mechanics and Physics of Solids 19~(4) (1971) 179--188.

\bibitem{Sun2021d}
W.~Sun, S.~Wu, {A study of crack initiation and source mechanism in the
  Brazilian test based on moment tensor}, Engineering Fracture Mechanics 246
  (2021) 107622.

\bibitem{Duevel1997}
B.~Duevel, B.~Haimson, {Mechanical characterization of pink Lac du Bonnet
  granite: evidence on nonlinearity and anisotropy}, International journal of
  rock mechanics and mining sciences {\&} geomechanics abstracts 34~(3-4)
  (1997) 543.

\bibitem{Cai2010}
M.~Cai, {Practical estimates of tensile strength and Hoek-Brown strength
  parameter mi of brittle rocks}, Rock Mechanics and Rock Engineering 43~(2)
  (2010) 167--184.

\end{thebibliography}


\end{document}